\documentstyle[12pt,cite,epsfig]{article}
\skip\footins 10.8pt plus 4pt minus 2pt

\def\Thebibliography#1{\section*{REFERENCES}\list
 {[\arabic{enumi}]}{\settowidth\labelwidth{[#1]}\leftmargin\labelwidth
 \advance\leftmargin\labelsep
 \usecounter{enumi}}
 \def\newblock{\hskip .11em plus .33em minus .07em}
 \sloppy\clubpenalty4000\widowpenalty4000
 \sfcode`\.=1000\relax}

\newcommand{\be}{\begin{equation}}
\newcommand{\ee}{\end{equation}}
\newcommand{\ba}{\begin{array}{c}}
\newcommand{\ea}{\end{array}}
\newcommand{\beqn}{\begin{eqnarray}}
\newcommand{\eeqn}{\end{eqnarray}}
\newcommand{\dis}{\displaystyle}
\newcommand{\bi}{\begin{itemize}}
\newcommand{\ei}{\end{itemize}}
\newcommand{\lrder}{\stackrel{\leftrightarrow}{\partial}}
\newcommand{\cL}{{\cal L}}
\newcommand{\cD}{{\cal D}}
\newcommand{\cM}{{\cal M}}

\newcommand{\cP}{{\cal P}}

\newcommand{\dg}{\dagger}
\newcommand{\dfrac}{\displaystyle \frac}
\newcommand{\no}{\nonumber}
\newcommand{\lsim}{\stackrel{<}{_\sim}}
\newcommand{\rms}{\rm\scriptsize}
\begin{document}

%
%

\begin{titlepage}
\vspace{0.3in}
\begin{flushright}
{CERN-TH.6978/93}
\end{flushright}
\vspace*{1.2cm}
\begin{center}
{\Large \bf INTRODUCTION TO \\ CHIRAL PERTURBATION THEORY}

\vspace*{0.4cm}
A. Pich$^{*\dagger}$

Theory Division, CERN, CH-1211 Geneva 23

\end{center}
\vspace*{1.5cm}

\begin{abstract}
\noindent An introduction to the basic ideas and methods of
Chiral Perturbation Theory is presented.
Several phenomenological applications of the
effective Lagrangian technique to strong, electromagnetic
and weak interactions are discussed.
\end{abstract}

\vspace*{2.2cm}
\begin{center}
{Lectures given at the\\
V Mexican School of Particles and Fields \\
Guanajuato, M\'exico, December 1992}
\end{center}

\vspace*{2.5cm}

{\footnotesize\baselineskip 10pt \noindent
$^*$ On leave of absence from
Departament de F\'{\i}sica Te\`orica, Universitat de Val\`encia,
and IFIC, Centre Mixte Universitat de Val\`encia--CSIC,
E-46100 Burjassot, Val\`encia, Spain.}

{\footnotesize\baselineskip 10pt \noindent
$^\dagger$ Work supported in part
by CICYT (Spain), under grant No. AEN90-0040.}

\vfill
\begin{flushleft}
{CERN-TH.6978/93 \\
August 1993}
\end{flushleft}
\end{titlepage}

%

\title{INTRODUCTION TO \\ CHIRAL PERTURBATION THEORY}
\author{A. Pich\thanks{On leave of absence from
Departament de F\'{\i}sica Te\`orica, Universitat de Val\`encia,
and IFIC, Centre Mixte Universitat de Val\`encia--CSIC,
E-46100 Burjassot, Val\`encia, Spain.}\ \thanks{Work supported in part
by CICYT (Spain), under grant No. AEN90-0040.}
 \\ Theory Division, CERN, CH-1211 Geneva 23}\date{}
\maketitle
\begin{abstract}
An introduction to the basic ideas and methods of
Chiral Perturbation Theory is presented.
Several phenomenological applications of the
effective Lagrangian technique to strong, electromagnetic
and weak interactions are discussed.
\end{abstract}
\section{Effective Field Theories}
\label{sec:introduction}

Effective Field Theories (EFTs)
are the appropriate theoretical tool to describe
``low-energy'' physics, where ``low'' is defined with respect to some
energy scale $\Lambda$.
What that means is that they only take explicitly into account
the relevant
degrees of freedom, i.e. those states with $m<<\Lambda$, while the
heavier excitations with $M>>\Lambda$ are integrated out from the action.
One gets in this way a string of non-renormalizable interactions
among the light states, which can be organized as an expansion
in powers of energy/$\Lambda$.
The information on  the heavier degrees of freedom is then
contained in the couplings of the resulting low-energy Lagrangian.
Although EFTs contain an infinite number of terms, renormalizability
is not an issue since, at a given order in the energy expansion,
the low-energy theory is specified by a finite number of couplings;
this allows for an order-by-order renormalization.
Obviously, for this procedure to make sense, it is necessary that the
spectrum of the fundamental theory contains a mass gap, separating
the light and heavy states.

A simple example of EFT is provided by QED at very low energies,
$\omega<<m_e$, where $\omega$ denotes the photon energy.
In this limit, one can describe the light-by-light scattering
using an  effective Lagrangian in terms of the electromagnetic field only.
Gauge and Lorentz invariance
constrain the possible structures present
in the effective Lagrangian:
\be\label{eq:QED}
{\cL}_{\mbox{\rms eff}} = -{1\over 4} F^{\mu\nu} F_{\mu\nu}
   + {a\over m_e^4} \, (F^{\mu\nu} F_{\mu\nu})^2
   + {b\over m_e^4} \, F^{\mu\nu} F_{\nu\sigma}  F^{\sigma\rho} F_{\rho\mu}
   + O(F^6/m_e^8) .
\ee
In the low-energy regime, all the information on the original QED dynamics
is embodied in the values of the two low-energy couplings $a$ and $b$.
The values of these constants can be computed,
by explicitly integrating out the electron field from the original
QED generating
functional (or equivalently, by computing the relevant light-by-light
box diagrams). One then gets the well-known Euler-Heisenberg result
\cite{ref:EH}:
\be\label{eq:EH}
a = -{\alpha^2\over 36}, \qquad\qquad\qquad
b = {7 \alpha^2\over 90} .
\ee
The important point to realize is that, even in the absence of an
explicit computation of the couplings $a$ and $b$, the Lagrangian
(\ref{eq:QED}) contains non-trivial information, which is a consequence
of the imposed symmetries. The dominant contributions to the
amplitudes for different
low-energy photon reactions like $\gamma\gamma\to 2\gamma, 4\gamma, \ldots$
can be directly obtained from
${\cL}_{\mbox{\rms eff}}$.
Moreover, the order of magnitude of the constants $a$, $b$ can also be
easily estimated through a na\"{\i}ve counting of powers of the
electromagnetic coupling and combinatorial and loop
[$1/(16 \pi^2)$] factors.

The previous example is somehow academic, since perturbation theory
in powers of $\alpha$ works extremely well in QED. However, the effective
Lagrangian (\ref{eq:QED}) would be valid even if the fine structure
constant were big; the only difference would then be that we would not be
able to perturbatively compute the couplings $a$ and $b$.

We can mention two generic situations where
EFTs become particularly useful:

\bi

\item The underlying fundamental theory is unknown, but the symmetry
properties of the light states can be used to build an effective Lagrangian.
The low-energy couplings then parametrize the unknown new physics.
A typical example
are EFTs at the electroweak scale.

\item Even if the underlying fundamental theory is known, sometimes it is
not directly applicable in the low-energy region. For instance,
due to confinement, the
quark and gluon fields of QCD are not asymptotic states.
Since we do not know how to solve QCD, we cannot derive the hadronic
interactions directly from the original QCD Lagrangian. However, we
do know the symmetry properties of the strong interactions; therefore,
we can write an EFT in terms of the hadronic asymptotic states, and
parametrize the unknown dynamical information in a few couplings.

\ei

The theoretical basis of EFTs can be formulated \cite{ref:WE79}
as a ``theorem''\footnote{
Although this ``theorem'' is almost self-evident, it is only ``proven''
to the extent that no counter-examples are known.}:
for a given set of asymptotic states, perturbation theory with the
most general Lagrangian containing all terms allowed by the assumed
symmetries will yield the most general S-matrix elements consistent
with analyticity, perturbative unitarity
and the assumed symmetries.

The purpose of these lectures is to give a pedagogical
introduction to Chiral Perturbation Theory (ChPT), the
low-energy effective field theory of the Standard Model.
The chiral symmetry of the QCD Lagrangian is discussed
in Section~\ref{sec:symmetry}, and a toy low-energy model
incorporating the right symmetry properties is studied
in Section~\ref{sec:sigma}.
The ChPT formalism is presented in
Sections~\ref{sec:lo} and \ref{sec:p4},
where
the lowest-order and next-to-leading-order
terms in the chiral expansion
are analysed.
Section~\ref{sec:phenomenology} contains a few selected
phenomenological applications.
The relation between the effective Lagrangian and
the underlying fundamental QCD theory is studied
in Section~\ref{sec:couplings}, which summarizes
recent attempts to calculate the chiral couplings.
The effective realization of the non-leptonic
$\Delta S=1$ interactions is described in
Section~\ref{sec:weak}, and a brief overview of
the application of the chiral techniques to
$K$ decays is given in
Sections~\ref{sec:kpp}, \ref{sec:radiative} and
\ref{sec:anomalous}.
Section~\ref{sec:light_Higgs} shows how ChPT can be used
to work out the low-energy interactions of a
possible light Higgs boson.
Finally, Section~\ref{ref:electroweak}
illustrates the use of the chiral techniques
to describe the Goldstone dynamics associated with
the Standard Model electroweak symmetry breaking.
A few summarizing comments are collected in Section~\ref{sec:summary}.

To prepare these lectures I have made extensive use
of excellent reviews
\cite{ref:BI93,ref:DO92,ref:EC92,ref:GA90,ref:LE91,ref:ME93,ref:PI91,ref:RA89}
and books \cite{ref:RINBERG88,ref:DOBOGOKO91,ref:GE84,ref:DGH92}
already existing in the literature.
In many cases, I have sacrificed some rigour to simplify
the presentation of the subject.
A more careful discussion and
further details can be found in those
references.

\section{Chiral Symmetry}
\label{sec:symmetry}

   In the absence of quark masses, the QCD Lagrangian
[$q = \hbox{\rm column}(u,d,\ldots)$]
\be\label{eq:LQCD}
{\cal L}_{QCD}^0 = -{1\over 4} \mbox{\rm Tr}(G_{\mu\nu} G^{\mu\nu})
 + i \bar q_L \gamma^\mu D_\mu q_L  + i \bar q_R \gamma^\mu D_\mu q_R
\ee
 is invariant under
independent global $G\equiv SU(N_f)_L\otimes SU(N_f)_R$
transformations\footnote{
Actually, the Lagrangian (\protect\ref{eq:LQCD})
has a larger $U(N_f)_L\otimes U(N_f)_R$ global symmetry. However, the
$U(1)_A$ part is broken by quantum effects [$U(1)_A$ anomaly],
while the
quark-number symmetry $U(1)_V$ is trivially realized in the meson sector.
A discussion of the $U(1)_A$ part, within ChPT, is given in
refs.~\protect\cite{ref:GL85,ref:PR91b}.}
 of the left- and
right-handed quarks in flavour space:
\be\label{eq:qrot}
 q_L \, \stackrel{G}{\longrightarrow} \, g_L \, q_L , \qquad\qquad
q_R \, \stackrel{G}{\longrightarrow} \, g_R \, q_R , \qquad\qquad
g_{L,R} \in SU(N_f)_{L,R} .
\ee
The Noether currents associated with the chiral group $G$ are
[$\lambda_a$ are Gell-Mann's matrices with
$\hbox{\rm Tr}(\lambda_a\lambda_b) = 2 \delta_{ab}$]:
\be
J^{a\mu}_X = \bar q_X \gamma^\mu {\lambda_a\over 2} q_X ,
\qquad\qquad (X = L,R;\quad a = 1,\,\ldots,\, 8) .
\ee
The corresponding Noether charges
$Q^a_X = \int d^3x J^{a0}_X(x)$ satisfy the familiar commutation relations
\be
[Q_X^a,Q_Y^b] = i \delta_{XY} f_{abc} Q^c_X ,
\ee
which were the starting point of the Current Algebra
methods of the sixties \cite{ref:currentalgebra}.

This chiral symmetry, which should be
 approximately good in the light quark
sector ($u$,$d$,$s$), is however not seen in
the hadronic spectrum. Although hadrons can be nicely classified in
$SU(3)_V$ representations,
 degenerate multiplets with opposite parity do not exist.
Moreover, the octet of pseudoscalar mesons happens to be much lighter than
all the other hadronic states.
  To be consistent with this experimental  fact, the ground state of the
theory (the vacuum) should not be symmetric under the chiral group.
The $SU(3)_L \otimes SU(3)_R$ symmetry
spontaneously breaks down to
$SU(3)_{L+R}$
and, according to Goldstone's theorem \cite{ref:GO61},
an octet of pseudoscalar massless
bosons appears in the theory.

More specifically,
let us consider a Noether charge $Q$,
and assume the existence of an
operator $O$ that satisfies
\be\label{eq:order}
\langle 0 | [Q,O] | 0 \rangle \not= 0 ;
\ee
this is clearly only possible if $Q|0\rangle\not= 0$.
Goldstone's theorem then tells us that there exists a
massless state $|G\rangle$
such that
\be
\langle 0|J^0|G\rangle \, \langle G|O|0\rangle\not= 0.
\ee
The quantum numbers of the Goldstone boson are
dictated by those of $J^0$ and $O$.
The quantity in the left-hand side of Eq. (\ref{eq:order})
is called the order parameter of the spontaneous symmetry breakdown.

Since there are eight broken axial generators of the chiral
group, $Q^a_A = Q^a_R - Q^a_L$,
there should be eight pseudoscalar Goldstone states
$|G^a\rangle$, which we can identify with
the eight lightest hadronic states
($\pi^+$, $\pi^-$, $\pi^0$, $\eta$, $K^+$, $K^-$, $K^0$ and $\bar{K}^0$);
their small masses being generated by the quark-mass matrix, which explicitly
breaks the global symmetry of the QCD Lagrangian.
The corresponding $O^a$ must be pseudoscalar operators. The simplest
possibility are $O^a = \bar q \gamma_5 \lambda_a q$, which satisfy
\be
\langle 0|[Q^a_A, \bar q \gamma_5 \lambda_b q] |0\rangle=
-{1\over 2} \,\langle 0|\bar q \{\lambda_a,\lambda_b\} q |0\rangle =
-{2\over 3} \,\delta_{ab} \,\langle 0|\bar q q |0\rangle .
\ee
The quark condensate
\be
\langle 0|\bar u u |0\rangle =
\langle 0|\bar d d |0\rangle =
\langle 0|\bar s s |0\rangle \not = 0
\ee
is then
the natural-order parameter of
Spontaneous Chiral Symmetry Breaking (SCSB).

The Goldstone nature of the pseudoscalar mesons implies strong
constraints on their interactions, which can be most
easily analysed on the
basis of an effective Lagrangian.

\section{A toy Lagrangian: the linear sigma model}
\label{sec:sigma}

The linear sigma model \cite{ref:sigma}
\beqn\label{eq:sigma}
{\cal L}_\sigma & = & {1\over 2}
\left[ \partial_\mu\sigma\partial^\mu\sigma
+ \partial_\mu\vec{\pi}\partial^\mu\vec{\pi} \right]
- V(\sigma,\vec{\pi}) , \\ \label{eq:potential}
V(\sigma,\vec{\pi}) & = & {\lambda\over 4} \left( \sigma^2 + \vec{\pi}^2
- v^2 \right)^2 , \qquad \qquad (\lambda>0), \no
\eeqn
provides a very simple example of SCSB.
If $v^2<0$,
the global symmetry $O(4)\sim SU(2)\otimes SU(2)$ is realized in the
usual Wigner--Weyl way; one then has degenerate $\sigma,\vec{\pi}$
states with mass $m^2=-\lambda v^2$.
However, for $v^2>0$, the
potential $V(\sigma,\vec{\pi})$
has a  family of minima occurring for all $\sigma,\vec{\pi}$
with $\sigma^2 + \vec{\pi}^2 = v^2$; these minima correspond to
degenerate ground states, which transform into each other under
chiral rotations. The symmetry is then realized \`a la Nambu--Goldstone,
and three massless states appear, corresponding to the
flat directions of $V(\sigma,\vec{\pi})$.
Taking
\be\label{eq:choice}
\langle 0|\sigma|0\rangle = v , \qquad
\langle 0|\vec{\pi}|0\rangle = 0 ,
\ee
and making the field redefinition $\hat{\sigma} = \sigma - v$,
the Lagrangian takes the form
\be\label{eq:sigma2}
{\cal L}_\sigma  =  {1\over 2}
\left[ \partial_\mu\hat{\sigma}\partial^\mu\hat{\sigma}
- 2 \lambda v^2 \hat{\sigma}^2
+ \partial_\mu\vec{\pi}\partial^\mu\vec{\pi} \right]
 - \lambda v \hat{\sigma} \left( \hat{\sigma}^2 + \vec{\pi}^2\right)
- {\lambda\over 4} \left( \hat{\sigma}^2 + \vec{\pi}^2\right)^2 ,
\ee
which  shows that  $\vec{\pi}$ corresponds to the three
massless Goldstone modes, while the $\hat{\sigma}$ field acquires
a mass $m_{\hat{\sigma}}^2 = 2 \lambda v^2$.

To clarify the role of chiral symmetry on the Goldstone dynamics,
it is useful to rewrite the sigma-model Lagrangian in a different
way.
Using the $2\times 2$ matrix notation
\be\label{eq:notation}
\Sigma(x) \equiv \sigma(x) I + i \vec{\tau}\vec{\pi} ,
\ee
the Lagrangian (\ref{eq:sigma}) takes the compact form
\be\label{eq:sigma3}
{\cal L}_\sigma = {1\over 4}
\langle \partial_\mu\Sigma^\dagger\partial^\mu\Sigma\rangle
- {\lambda \over 16}
\left( \langle \Sigma^\dagger\Sigma\rangle - 2 v^2 \right)^2 ,
\ee
where $\langle A \rangle$ denotes the trace of the matrix $A$.
In this notation the Lagrangian is explicitly invariant under
global chiral $G\equiv SU(2)_L\otimes SU(2)_R$ transformations:
\be
\Sigma \, \stackrel{G}{\longrightarrow} \,
g_R \,\Sigma\, g_L^\dagger , \qquad\qquad
g_{L,R}  \in SU(2)_{L,R} .
\ee
We can now make the polar decomposition
\cite{ref:EC92}
\beqn\label{eq:polar}
\Sigma(x) & = & \left( v + S(x) \right) \, U(\phi(x)) , \\
 U(\phi(x)) & = & \exp{\left( i \vec{\tau} \vec{\phi}(x) / v \right) } ,
\no
\eeqn
in terms of a Hermitian scalar field $S$ and
pseudoscalar fields $\vec{\phi}$.
These fields transform in a non-linear way under the chiral group:
\be\label{eq:transf}
S \, \stackrel{G}{\longrightarrow} \, S,
\qquad\qquad\qquad
 U(\phi) \, \stackrel{G}{\longrightarrow} \, g_R\, U(\phi)\, g_L^\dagger .
\ee
The sigma-model Lagrangian then takes the form
\be\label{eq:sigma4}
{\cal L}_\sigma  =  {v^2\over 4} \left ( 1 + {S\over v} \right)^2
\langle \partial_\mu U^\dagger \partial^\mu U \rangle
 + {1\over 2}
\left( \partial_\mu S \partial^\mu S - 2 \lambda v^2 S^2\right)
- \lambda v S^3 - {\lambda\over 4} S^4 .
\ee
\goodbreak

Equation (\ref{eq:sigma4}) is very instructive:
\bi
\item It shows explicitly that the Goldstone bosons have purely
derivative couplings, as they should.
This was not so obvious in Eq. (\ref{eq:sigma2}). Of course one should
get the same measurable amplitudes from both Lagrangians, which
means that the original Lagrangian  (\ref{eq:sigma2}) gives rise to
exact (and not very transparent) cancellations among different
momentum-independent contributions.
\item In the limit $\lambda >> 1$, the scalar field $S$ becomes very
heavy and can be integrated out from the Lagrangian. The linear sigma
model then reduces to the familiar Lagrangian
\be\label{eq:universal}
{\cal L}_2 = {v^2\over 4}
\langle \partial_\mu U^\dagger \partial^\mu U \rangle .
\ee
As we will see in the next section, this is a universal
model-independent interaction
of the Goldstone bosons induced by SCSB.
It is often claimed in the literature that the linear sigma model
correctly describes the low-energy strong interactions.
This is, however,
quite a misleading statement. To the extent that one is only looking
at the
predictions coming from the model-independent
lowest-order term (\ref{eq:universal}), the comparison with experiment
only tests the assumed pattern of SCSB.

\item In order to be sensitive to the particular structure
of the linear sigma model, one needs to test the model-dependent part
involving the scalar field $S$. At low momenta ($p << M_S$), the
dominant tree-level corrections originate from the exchange
of an $S$ particle, which generates the four-derivative term
\be\label{eq:sigma5}
{\cal L}_\sigma^4 = {v^2\over 8 M_S^2}
\langle \partial_\mu U^\dagger \partial^\mu U \rangle^2 .
\ee
It will be shown later that this kind of
interaction does not agree with the
experimental data. Therefore, the linear sigma model is not
a phenomenologically viable EFT of QCD \cite{ref:GL84}.

\ei

\section{Effective Chiral Lagrangian at lowest order}
\label{sec:lo}

We want to get an effective Lagrangian realization of QCD, at low
energies, for the light-quark sector ($u$, $d$, $s$).
Our basic assumption is the pattern of SCSB:
\be
G \equiv SU(3)_L\otimes SU(3)_R \stackrel{SCSB}{\longrightarrow}
H \equiv SU(3)_V .\ee
The present understanding of the mechanism of SCSB
is based in the dynamical generation of a non-zero vacuum expectation
value of the scalar quark density, i.e. the vacuum condensate
$v \equiv \langle 0| \bar u u |0\rangle =
\langle 0| \bar d d |0\rangle =
\langle 0| \bar s s |0\rangle \not= 0$.
The Goldstone bosons correspond to the zero-energy excitations over
this vacuum condensate; their fields can be collected in a $3\times 3$
unitary matrix $U(\phi)$,
\be
\langle 0| \bar q^j_L q^i_R|0\rangle \, \longrightarrow \, {v\over 2}\,
U^{ij}(\phi) ,
\ee
which parametrizes those excitations.
A convenient parametrization is given by
\be
U(\phi)  \equiv \exp{\left(i \sqrt{2} \Phi / f\right)} ,
\ee
where
%
\be
\ba
\Phi (x) \equiv
\frac{\dis \vec{\lambda}}{\dis \sqrt 2} \, \vec{\phi}
 = \, \left( \begin{array}{ccc}
\frac{\dis \pi^0}{\dis \sqrt 2} \, + \, \frac{\dis \eta_8}{\dis \sqrt 6}
 & \pi^+ & K^+ \\
\pi^- & - \frac{\dis \pi^0}{\dis \sqrt 2} \, + \, \frac{\dis \eta_8}
{\dis \sqrt 6}    & K^0 \\
K^- & \bar K^0 & - \frac{\dis 2 \, \eta_8}{\dis \sqrt 6}
\end{array}  \right) .
\ea
\ee
The matrix
$U(\phi)$ transforms linearly under the chiral group,
\be\label{eq:utransf}
U(\phi) \, \stackrel{G}{\longrightarrow}\, g_R \, U(\phi) \, g_L^\dagger ,
\ee
but the induced transformation on the Goldstone fields
$\vec{\phi}$ is highly non-linear.

Since there is a mass gap separating the pseudoscalar octet
from the rest of the hadronic spectrum,
we can build an EFT containing only the Goldstone modes.
We should write the more general Lagrangian involving the matrix
$U(\phi)$, which is consistent with chiral symmetry.
Moreover, we can
organize the Lagrangian in terms of increasing powers of
momentum or, equivalently, in terms of an increasing number of
derivatives (parity conservation requires an even number of derivatives):
\be
{\cal L}_{\hbox{\rms eff}}(U) \, = \, \sum_n {\cal L}_{2n} .
\ee
 In the low-energy domain we are interested in, the
terms with a minimum number of derivatives will dominate.

Due to the unitarity of the $U$ matrix, $U U^\dagger = I$, at least
two derivatives are required to generate a non-trivial interaction.
To lowest order, the effective chiral Lagrangian is uniquely
given by the term
\be\label{eq:l2}
{\cal L}_2 = {f^2\over 4}
\langle \partial_\mu U^\dagger \partial^\mu U \rangle .
\ee
This is exactly the structure (\ref{eq:universal}), which we
derived from the linear sigma model in the last section.

Expanding $U(\phi)$ in a power series in $\phi$, one obtains the
Goldstone's kinetic terms plus a tower of interactions involving
an increasing number of pseudoscalars.
The requirement that the kinetic terms are properly normalized
fixes the global coefficient $f^2/4$ in Eq. (\ref{eq:l2}).
All interactions among the Goldstones can then be predicted in terms
of the single coupling $f$:
\be
{\cal L}_2 \, = \, {1\over 2} \,\langle\partial_\mu\phi
\partial^\mu\phi\rangle
\, + \, {1\over 12 f^2} \,\langle (\phi\lrder_\mu\phi) \,
(\phi\stackrel{\leftrightarrow}{\partial^\mu}\phi)
\rangle \, + \, O(\phi^6/f^4) .
\ee

To compute the $\pi\pi$ scattering amplitude, for instance, is now
a trivial perturbative exercise. One gets the well-known \cite{ref:WE66}
Weinberg result  [$t\equiv (p_+' - p_+)^2$]
\be\label{eq:WE1}
T(\pi^+\pi^0\to\pi^+\pi^0) = {t\over f^2}.
\ee
Similar results can be obtained for $\pi\pi\to 4\pi, 6\pi, 8\pi, \ldots$\,
It is the non-linearity of the effective Lagrangian that  relates
amplitudes with different numbers of Goldstone bosons, allowing
for absolute predictions in terms of $f$.

The EFT technique becomes much more powerful if one introduces couplings
to external classical fields.
Let us consider an extended QCD Lagrangian, with quark
couplings to external Hermitian matrix-valued fields
$v_\mu$, $a_\mu$, $s$, $p$:
\be\label{eq:extendedqcd}
{\cal L}_{QCD} = {\cal L}^0_{QCD} +
\bar q \gamma^\mu (v_\mu + \gamma_5 a_\mu ) q -
\bar q (s - i \gamma_5 p) q .
\ee
The external fields will allow us to compute the effective realization
of general Green functions of quark currents in a very straightforward
way. Moreover, they can be used to incorporate the
electromagnetic and semileptonic weak interactions, and the
explicit breaking of chiral symmetry through the quark masses:
\beqn\label{eq:breaking}
r_\mu & \equiv & v_\mu + a_\mu \, = \, e Q A_\mu + \ldots
\nonumber\\
\ell_\mu & \equiv & v_\mu - a_\mu \, =
\,  e Q A_\mu + {e\over\sqrt{2}\sin{\theta_W}}
(W_\mu^\dagger T_+ + \mbox{\rm h.c.}) + \ldots
\\
s & = & {\cal M} + \ldots \nonumber
\eeqn
Here, $Q$ and ${\cal M}$ denote the quark-charge and quark-mass matrices,
respectively,
\be
Q = {1\over 3}\, \hbox{\rm diag}(2,-1,-1), \qquad\qquad
{\cal M} = \hbox{\rm diag}(m_u,m_d,m_s) ,
\ee
and $T_+$ is a $3\times 3$ matrix containing the relevant
Cabibbo--Kobayashi--Maskawa factors
\be
T_+ \, = \, \left(
\begin{array}{ccc}
0 & V_{ud} & V_{us} \\ 0 & 0 & 0 \\ 0 & 0 & 0
\end{array} \right).
\ee

Formally, the Lagrangian (\ref{eq:extendedqcd})
is invariant under the following set of local
$SU(3)_L\otimes SU(3)_R$ transformations:
\beqn\label{eq:symmetry}
q_L & \longrightarrow & g_L \, q_L , \nonumber\\
q_R & \longrightarrow & g_R \, q_R , \nonumber\\
\ell_\mu & \longrightarrow & g_L \, \ell_\mu \, g_L^\dagger \, + \,
i g_L \partial_\mu g_L^\dagger ,
\\
r_\mu & \longrightarrow & g_R \, r_\mu \, g_R^\dagger \, + \,
i g_R \partial_\mu g_R^\dagger ,
\nonumber\\
s + i p & \longrightarrow & g_R \, (s + i p) \, g_L^\dagger .
\nonumber
\eeqn
We can use this formal symmetry to build a generalized effective
Lagrangian for the Goldstone bosons, in the presence of external
sources. Note that to respect the local invariance, the gauge fields
$v_\mu$, $a_\mu$ can only appear through the covariant derivative
\be
D_\mu U = \partial_\mu U - i r_\mu U + i U \ell_\mu ,
\qquad
D_\mu U^\dagger = \partial_\mu U^\dagger  + i U^\dagger r_\mu
- i \ell_\mu U^\dagger ,
\ee
and through the field strength tensors
\be
F^{\mu\nu}_L =
\partial^\mu \ell^\nu - \partial^\nu \ell^\mu - i [ \ell^\mu , \ell^\nu ] ,
\qquad
F^{\mu\nu}_R =
\partial^\mu r^\nu - \partial^\nu r^\mu - i [ r^\mu , r^\nu ] .
\ee
At lowest order in momenta, the more general effective Lagrangian
consistent with Lorentz invariance and with (local) chiral symmetry
is of the form \cite{ref:GL85}
\be\label{eq:lowestorder}
{\cal L}_2 = {f^2\over 4}\,
\langle D_\mu U^\dagger D^\mu U \, + \, U^\dagger\chi  \, +  \,\chi^\dagger U
\rangle ,
\ee
where
\be
\chi \, = \, 2 B_0 \, (s + i p) ,
\ee
and $B_0$ is a constant, which, like $f$, is not fixed by
symmetry requirements alone.

Once special directions in flavour space, like the ones in
Eq. (\ref{eq:breaking}), are selected for the external fields,
chiral symmetry is of course explicitly broken.
The important point is that (\ref{eq:lowestorder}) then breaks the
symmetry in exactly the same way as the fundamental short-distance
Lagrangian (\ref{eq:extendedqcd}) does.

The power of the external field technique becomes obvious when
computing the chiral Noether currents.
Formally, the physical Green functions are obtained as functional
derivatives of the generating functional
$Z[v,a,s,p]$, defined via the path-integral formula
\be\label{eq:generatingfunctional}
\exp{\{i Z\}} \, = \, \int \, {\cal D}q \,{\cal D} \bar q
\,{\cal D}G_\mu \,
\exp{\left\{i \int d^4x\, {\cal L}_{QCD}\right\}} \, = \,
\int \, {\cal D}U(\phi) \,
\exp{\left\{i \int d^4x\, {\cal L}_{\mbox{\rms eff}}\right\}} .
\ee
At lowest order in momenta, the generating functional reduces to the
classical action $S_2 = \int d^4x \,{\cal L}_2$;
therefore, the currents can be trivially
computed by taking the appropriate derivatives
 with respect to the external fields:
\beqn
J^\mu_L \doteq {\delta S_2\over \delta \ell_\mu} & = &
 {i\over 2} f^2 D_\mu U^\dagger U =
\hphantom{-}{f\over\sqrt{2}} D_\mu \phi -
{i\over 2} \, \left(\phi
\stackrel{\leftrightarrow}{D^\mu}\phi\right) +
O(\phi^3/f) , \no\\
J^\mu_R \doteq {\delta S_2\over \delta r_\mu} & = &
 {i\over 2} f^2 D_\mu U U^\dagger =
-{f\over\sqrt{2}} D_\mu \phi -
{i\over 2} \, \left(\phi
\stackrel{\leftrightarrow}{D^\mu}\phi\right) +
O(\phi^3/f) .
\eeqn

The physical meaning of the chiral coupling $f$ is now obvious;
at $O(p^2)$, $f$ equals the pion decay constant,
$f = f_\pi = 93.2$ MeV, defined as
\be
\langle 0 | (J^\mu_A)^{12} | \pi^+\rangle
 \equiv i \sqrt{2} f_\pi p^\mu .
\ee
Similarly, by taking a derivative with respect to the external scalar
source $s$, we learn that the constant $B_0$ is related to the
quark condensate
\be\label{eq:b0}
\langle 0 | \bar q^j q^i|0\rangle = -f^2 B_0 \delta^{ij} .
\ee

Taking $s = {\cal M}$ and $p=0$,
the $\chi$ term in Eq. (\ref{eq:lowestorder})
gives rise to a quadratic pseudoscalar-mass term plus
additional interactions proportional to the quark masses.
Expanding in powers of $\phi$
(and dropping an irrelevant constant), one has
\be\label{eq:massterm}
{f^2\over 4} 2 B_0 \,\langle {\cal M} (U + U^\dagger) \rangle
=  B_0 \left\{ - \langle {\cal M}\phi^2\rangle
+ {1\over 6 f^2} \langle {\cal M} \phi^4\rangle
+ O(\phi^6/f^4) \right\} .
\ee

The explicit evaluation of the trace in the quadratic mass term provides
the relation between the physical meson masses and the quark masses:
\beqn\label{eq:masses}
M_{\pi^\pm}^2 & = & 2 \hat{m} B_0 , \nonumber\\
M_{\pi^0}^2 & = & 2 \hat{m} B_0 - \varepsilon +
O(\varepsilon^2) , \nonumber\\
M_{K^\pm}^2 & = & (m_u + m_s) B_0 , \\
M_{K^0}^2 & = & (m_d + m_s) B_0 , \nonumber\\
M_{\eta_8}^2 & = & {2\over 3} (\hat{m} + 2 m_s)  B_0 + \varepsilon +
O(\varepsilon^2) , \no
\eeqn
where\footnote{
The $O(\varepsilon)$ corrections to $M_{\pi^0}^2$
and $M_{\eta_8}^2$
originate from a small mixing term between the
$\pi^0$ and $\eta_8$ fields,
$$
- B_0 \langle {\cal M}\phi^2\rangle \longrightarrow
- (B_0/\sqrt{3})\, (m_u - m_d)\, \pi^0\eta_8 .
$$
The diagonalization of the quadratic $\pi^0$, $\eta_8$
mass matrix, gives
the mass eigenstates,
$\pi^0 = \cos{\delta} \,\phi^3 + \sin{\delta} \,\phi^8$
and
$\eta_8 = -\sin{\delta} \,\phi^3 + \cos{\delta} \,\phi^8$,
where
$
\tan{(2\delta)} = \sqrt{3} (m_d-m_u)/\left( 2 (m_s-\hat{m})\right) .
$
}
\be
\hat{m} = {1\over 2} (m_u + m_d) , \qquad\qquad
\varepsilon = {B_0\over 4} {(m_u - m_d)^2\over  (m_s - \hat{m})} .
\ee

Chiral symmetry relates the magnitude of the meson and quark masses
to the size of the quark condensate.
Using the result (\ref{eq:b0}), one gets from
the first equation in (\ref{eq:masses})
the well-known Gell-Mann--Oakes--Renner relation \cite{ref:GMOR}
\be
f^2_\pi M_\pi^2 = -  \hat{m} \,\langle 0|\bar u u + \bar d d|0\rangle .
\ee

Taking out the common $B_0$ factor, Eqs. (\ref{eq:masses}) imply
the old Current Algebra mass ratios \cite{ref:GMOR,ref:WE77},
\be\label{eq:mratios}
{M^2_{\pi^\pm}\over 2 \hat{m}} = {M^2_{K^+}\over (m_u+m_s)} =
{M_{K^0}\over (m_d+m_s)}
\approx {3 M^2_{\eta_8}\over (2 \hat{m} + 4 m_s)} ,
\ee
and
(up to $O(m_u-m_d)$ corrections)
the Gell-Mann--Okubo mass relation \cite{ref:GO},
\be
3 M^2_{\eta_8} = 4 M_K^2 - M_\pi^2 .
\ee
Note that the chiral Lagrangian 
automatically implies the successful quadratic
Gell-Mann--Okubo mass relation, and not a linear one.
Since $B_0 m_q\propto M^2_\phi$, the external field $\chi$
is counted as $O(p^2)$ in the chiral expansion.

Although chiral symmetry alone cannot fix the absolute values
of the quark masses, it gives information about quark-mass
ratios. Neglecting the tiny $O(\varepsilon)$ effects,
one gets the relations
\beqn\label{eq:ratio1}
{m_d - m_u \over m_d + m_u} & = &
{(M_{K^0}^2 - M_{K^+}^2) - (M_{\pi^0}^2 - M_{\pi^+}^2)\over M_{\pi^0}^2}
\, = \, 0.29 , \\ \label{eq:ratio2}
{m_s -\hat{m}\over 2 \hat{m}} & = &
{M_{K^0}^2 - M_{\pi^0}^2\over M_{\pi^0}^2}
\, = \, 12.6 .
\eeqn
In Eq. (\ref{eq:ratio1}) we have subtracted the pion square-mass
difference, to take into account the electromagnetic contribution
to the pseudoscalar-meson self-energies;
in the chiral limit ($m_u=m_d=m_s=0$), this contribution is proportional
to the square of the meson charge and it is the same for $K^+$ and $\pi^+$
\cite{ref:DA69}.
The mass formulae (\ref{eq:ratio1}) and (\ref{eq:ratio2})
imply the quark ratios advocated by Weinberg \cite{ref:WE77}:
\be\label{eq:Weinbergratios}
m_u : m_d : m_s = 0.55 : 1 : 20.3 .
\ee
Quark-mass corrections are therefore dominated by $m_s$, which is
large compared with $m_u$, $m_d$.
Notice that the difference $m_d-m_u$ is not small compared with
the individual up- and down-quark masses; in spite of that,
isospin turns out
to be an extremely good symmetry, because
isospin-breaking effects are governed by the small ratio
$(m_d-m_u)/m_s$.

The $\phi^4$ interactions in Eq. (\ref{eq:massterm})
introduce mass corrections to the $\pi\pi$ scattering amplitude
(\ref{eq:WE1}),
\be\label{eq:WE2}
T(\pi^+\pi^0\to\pi^+\pi^0) = {t - M_\pi^2\over f_\pi^2},
\ee
in perfect agreement with the Current Algebra result \cite{ref:WE66}.
Since $f\approx f_\pi$ is fixed from pion decay, this result
is now an absolute prediction of chiral symmetry!

The lowest-order chiral Lagrangian (\ref{eq:lowestorder}) encodes
in a very compact way all the Current Algebra results obtained in
the sixties \cite{ref:currentalgebra}.
The nice feature of the chiral approach is its elegant
simplicity. Moreover, as we will see in the next section, the EFT method
allows us to estimate higher-order corrections in a systematic way.

\section{ChPT at $O(p^4)$}
\label{sec:p4}

At next-to-leading order in momenta, $O(p^4)$, the
computation of the generating functional $Z[v,a,s,p]$ involves
three different ingredients:
\begin{enumerate}
\item The most general effective chiral Lagrangian of
$O(p^4)$, ${\cal L}_4$, to be considered at tree level.
\item One-loop graphs associated with the lowest-order
Lagrangian ${\cal L}_2$.
\item The Wess--Zumino--Witten functional \cite{ref:WZW}
to account for the chiral anomaly \cite{ref:anomaly,ref:BA69}.
\end{enumerate}

\subsection{$O(p^4)$ Lagrangian}

At $O(p^4)$, the most general\footnote{
Since we will only need ${\cal L}_4$ at tree level,
the general expression of this Lagrangian has been simplified,
using the $O(p^2)$ equations of motion obeyed by $U$.
Moreover, a $3\times 3$ matrix relation has been used to reduce the
number of independent terms.
For the two-flavour case, not all of these terms are independent
\protect\cite{ref:GL84,ref:GL85}.}
Lagrangian, invariant under
parity, charge conjugation and
the local chiral transformations (\ref{eq:symmetry}),
is given by \cite{ref:GL85}
\beqn\label{eq:l4}
{\cal L}_4 & = &
L_1 \,\langle D_\mu U^\dagger D^\mu U\rangle^2 \, + \,
L_2 \,\langle D_\mu U^\dagger D_\nu U\rangle\,
   \langle D^\mu U^\dagger D^\nu U\rangle
\nonumber\\  &  &
+~L_3 \,\langle D_\mu U^\dagger D^\mu U D_\nu U^\dagger
D^\nu U\rangle\,
+ \, L_4 \,\langle D_\mu U^\dagger D^\mu U\rangle\,
   \langle U^\dagger\chi +  \chi^\dagger U \rangle
\nonumber\\  &  &
+~L_5 \,\langle D_\mu U^\dagger D^\mu U \left( U^\dagger\chi +
\chi^\dagger U
\right)\rangle\,
+ \, L_6 \,\langle U^\dagger\chi +  \chi^\dagger U \rangle^2
\nonumber\\ &  &
+~L_7 \,\langle U^\dagger\chi -  \chi^\dagger U \rangle^2\,
+ \, L_8 \,\langle\chi^\dagger U \chi^\dagger U
+ U^\dagger\chi U^\dagger\chi\rangle
\nonumber\\  &  &
-~i L_9 \,\langle F_R^{\mu\nu} D_\mu U D_\nu U^\dagger +
     F_L^{\mu\nu} D_\mu U^\dagger D_\nu U\rangle\,
+ \, L_{10} \,\langle U^\dagger F_R^{\mu\nu} U F_{L\mu\nu} \rangle
\nonumber\\ &  &
+~H_1 \,\langle F_{R\mu\nu} F_R^{\mu\nu} +
F_{L\mu\nu} F_L^{\mu\nu}\rangle\,
+ \, H_2 \,\langle \chi^\dagger\chi\rangle .
\eeqn

The terms proportional to $H_1$ and $H_2$ do not contain the
pseudoscalar fields and are therefore not directly measurable.
Thus, at $O(p^4)$ we need ten additional coupling constants
$L_i$
to determine the low-energy behaviour of the Green functions.
These constants  parametrize our
ignorance about the details of the underlying QCD dynamics.
In principle, all the chiral couplings are calculable functions
of $\Lambda_{QCD}$ and the heavy-quark masses. At the present time,
however, our main source of information about these couplings
is low-energy phenomenology.

\subsection{Chiral loops}

ChPT is a quantum field theory, perfectly defined through
Eq. (\ref{eq:generatingfunctional}). As such, we must take
into account quantum loops with Goldstone-boson propagators in the
internal lines. The chiral loops generate non-polynomial contributions,
with logarithms and threshold factors, as required by unitarity.

The loop integrals are homogeneous functions of the external momenta and
the pseudoscalar masses occurring in the propagators.
A simple dimensional counting shows that,
for a general connected diagram with $N_d$ vertices of
$O(p^d)$ ($d=2,4,\ldots$),
$L$ loops and $I$ internal lines,
the overall
chiral dimension is given by \cite{ref:WE79}
\be
D = 2 L + 2 + \sum_d N_d (d-2) .
\ee
Each loop  adds two powers of momenta;
this power suppression of loop diagrams is at the basis of low-energy
expansions, such as ChPT.
The leading $D=2$ contributions are obtained with $L=0$ and $d=2$,
i.e. only tree-level graphs with ${\cal L}_2$ insertions.
At $O(p^4)$, we  have tree-level contributions from
${\cal L}_4$ ($L=0$, $d=4$, $N_4=1$) and one-loop graphs with the
lowest-order Lagrangian ${\cal L}_2$ ($L=1$, $d=2$).

ChPT is an expansion in powers of momenta over some typical hadronic
scale, usually called the scale of chiral symmetry breaking
$\Lambda_\chi$.
Since each chiral loop generates a geometrical factor
$(4 \pi)^{-2}$, plus a factor of $1/f^2$ to compensate
the additional dimensions,
one could expect \cite{ref:GM84}
$\Lambda_\chi$ to be about $4\pi f_\pi\sim 1.2 \, \mbox{\rm GeV}$.

The Goldstone loops are divergent and need to be renormalized.
Although EFTs are non-renormalizable (i.e. an infinite number
of counter-terms is required), order by order in the momentum expansion
they define a perfectly renormalizable theory.
If we use a regularization which preserves the symmetries of
the Lagrangian, such as dimensional regularization,
the counter-terms needed to renormalize the theory will be
necessarily symmetric.
Since by construction the full effective Lagrangian
contains all terms permitted by the symmetry,
the divergences can then be absorbed in a renormalization of the
coupling constants occurring in the Lagrangian.
At one loop (in ${\cal L}_2$), the ChPT divergences are $O(p^4)$
and are therefore renormalized by the low-energy couplings
in Eq. (\ref{eq:l4}):
\be\label{eq:renormalization}
L_i = L_i^r(\mu) + \Gamma_i \lambda , \qquad\qquad
H_i = H_i^r(\mu) + \widetilde\Gamma_i \lambda ,
\ee
where
\be
\lambda = {\mu^{d-4}\over 16 \pi^2} \left\{
{1\over d-4} -{1\over 2} \left[ \log{(4\pi)} + \Gamma'(1) + 1 \right]
\right\} .
\ee
The explicit calculation of the one-loop generating functional $Z_4$
\cite{ref:GL85} gives:
\beqn
\Gamma_1 = {3\over 32}, & \Gamma_2 = \dfrac{3}{16},\; & \Gamma_3 = 0\,\, ,
\qquad \Gamma_4 = {1\over 8} ,
\nonumber\\
\Gamma_5 = {3\over 8}\,\: , & \Gamma_6 = \dfrac{11}{144} , & \Gamma_7 = 0\,\, ,
\qquad \Gamma_8 = {5\over 48},
\\
\Gamma_9 = {1\over 4}\,\: , & \Gamma_{10} = -\dfrac{1}{4} , &
\widetilde\Gamma_1 = -{1\over 8}, \quad\,\, \widetilde\Gamma_2 = {5\over 24} .
\no
\eeqn
The renormalized couplings $L_i^r(\mu)$ depend on the arbitrary
scale of dimensional regularization $\mu$.
This scale dependence is of course
cancelled by that of the loop amplitude, in any physical,
measurable quantity.

A typical $O(p^4)$ amplitude will then consist of a non-polynomial
part, coming from the loop computation, plus a polynomial in momenta and
pseudoscalar masses, which depends on the unknown constants $L_i$.
The non-polynomial part (the so-called chiral logarithms) is
completely predicted as a function
of the lowest-order coupling $f$ and
the Goldstone masses.
This chiral structure can be easily understood in terms
of dispersion relations.
Given the lowest-order Lagrangian ${\cal L}_2$, the non-trivial
analytic behaviour associated with  some physical intermediate state
is calculable without the introduction of new arbitrary
chiral coefficients.
Analiticity then allows us to reconstruct the full amplitude, through
a dispersive integral, up to a subtraction polynomial.
ChPT generates (perturbatively) the correct dispersion integrals and
organizes the subtraction polynomials in a derivative expansion.

\subsection{The chiral anomaly}

Although the QCD Lagrangian (\ref{eq:extendedqcd})
is formally invariant under local
chiral transformations, this is no longer true for the associated
generating functional.
The anomalies of the fermionic determinant break chiral symmetry
at the quantum level \cite{ref:anomaly,ref:BA69}.
The anomalous change of the generating functional
under an infinitesimal chiral transformation
\be
g_{L,R} = 1 + i \alpha \mp i \beta + \ldots
\ee
%
is given by \cite{ref:BA69}:
\beqn\label{eq:anomaly}
&&\delta Z[v,a,s,p]  \, = \,
-{N_C\over 16\pi^2} \, \int d^4x \,
\langle \beta(x) \,\Omega(x)\rangle ,
\\
&&\Omega(x) \, = \,\varepsilon^{\mu\nu\sigma\rho} \,
 \left[
v_{\mu\nu} v_{\sigma\rho}
+ {4\over 3} \,\nabla_\mu a_\nu \nabla_\sigma a_\rho
+ {2\over 3} i \,\{ v_{\mu\nu},a_\sigma a_\rho\}
\right.\nonumber\\ && \qquad\qquad\qquad\left.\mbox{}
+ {8\over 3} i \, a_\sigma v_{\mu\nu} a_\rho
+ {4\over 3} \, a_\mu a_\nu a_\sigma a_\rho \right] ,
\no\\
&&v_{\mu\nu} \, = \,
\partial_\mu v_\nu - \partial_\nu v_\mu - i \, [v_\mu,v_\nu] ,
\qquad\qquad
\nabla_\mu a_\nu  \, = \,
\partial_\mu a_\nu - i \, [v_\mu,a_\nu] . \no
\eeqn
($N_C =3$ is the number of colours, and $\varepsilon_{0123}=1$.)
This anomalous variation of $Z$ is an $O(p^4)$
effect, in the chiral counting.

So far, we have been imposing chiral symmetry to construct the
effective ChPT Lagrangian.
Since chiral symmetry is explicitly violated
by the anomaly
at the fundamental
QCD level,
we need to add a functional $Z_A$ with the property that its
change under a chiral gauge transformation reproduces (\ref{eq:anomaly}).
Such a functional was constructed by Wess and Zumino \cite{ref:WZ71},
and reformulated in a nice geometrical way by Witten \cite{ref:WI83}.
It has the explicit form:
\beqn\label{eq:WZW}
S[U,\ell,r]_{WZW} &=&-\,\dfrac{i N_C}{240 \pi^2}
\int d\sigma^{ijklm} \left\langle \Sigma^L_i
\Sigma^L_j \Sigma^L_k \Sigma^L_l \Sigma^L_m \right\rangle
\no\\*
 & & -\,\dfrac{i N_C}{48 \pi^2} \int d^4 x\,
\varepsilon_{\mu \nu \alpha \beta}\left( W (U,\ell,r)^{\mu \nu
\alpha \beta} - W ({\bf 1},\ell,r)^{\mu \nu \alpha \beta} \right) ,
\\
W (U,\ell,r)_{\mu \nu \alpha \beta} & = &
\left\langle U \ell_{\mu} \ell_{\nu} \ell_{\alpha}U^{\dg} r_{\beta}
+ \frac{1}{4} U \ell_{\mu} U^{\dg} r_{\nu} U \ell_\alpha U^{\dg} r_{\beta}
+ i U \partial_{\mu} \ell_{\nu} \ell_{\alpha} U^{\dg} r_{\beta}
\right.\no  \\
& & +~ i \partial_{\mu} r_{\nu} U \ell_{\alpha} U^{\dg} r_{\beta}
- i \Sigma^L_{\mu} \ell_{\nu} U^{\dg} r_{\alpha} U \ell_{\beta}
+ \Sigma^L_{\mu} U^{\dg} \partial_{\nu} r_{\alpha} U \ell_\beta
\no \\
& & -~ \Sigma^L_{\mu} \Sigma^L_{\nu} U^{\dg} r_{\alpha} U \ell_{\beta}
+ \Sigma^L_{\mu} \ell_{\nu} \partial_{\alpha} \ell_{\beta}
+ \Sigma^L_{\mu} \partial_{\nu} \ell_{\alpha} \ell_{\beta}
\no\\
& & -~ i\left.  \Sigma^L_{\mu} \ell_{\nu} \ell_{\alpha} \ell_{\beta}
+ \frac{1}{2} \Sigma^L_{\mu} \ell_{\nu} \Sigma^L_{\alpha} \ell_{\beta}
- i \Sigma^L_{\mu} \Sigma^L_{\nu} \Sigma^L_{\alpha} \ell_{\beta}
\right\rangle \no \\
& & -~ \left( L \leftrightarrow R \right) ,
\eeqn
where
\be
\Sigma^L_\mu = U^{\dg} \partial_\mu U , \qquad\qquad
\Sigma^R_\mu = U \partial_\mu U^{\dg} ,
\ee
and
$\left( L \leftrightarrow R \right)$ stands for the interchanges
$U \leftrightarrow U^\dg $, $\ell_\mu \leftrightarrow r_\mu $ and
$\Sigma^L_\mu \leftrightarrow \Sigma^R_\mu $.
The integration in the first term of Eq. (\ref{eq:WZW}) is over a
five-dimensional manifold whose boundary is four-dimensional Minkowski
space. The integrand is a surface term; therefore both the first and the
second terms of $S_{WZW}$ are $O(p^4)$, according to the chiral
counting rules.

Since anomalies have a short-distance origin, their effect is
completely calculable. The translation from the fundamental
quark--gluon level to the effective chiral level is unaffected by
hadronization problems.
In spite of its considerable complexity, the anomalous action
(\ref{eq:WZW}) has no free parameters.

The anomaly functional gives rise to interactions that break
the intrinsic parity.
It is responsible for the $\pi^0\to 2\gamma$,
$\eta\to 2 \gamma$ decays, and the $\gamma 3\pi$, $\gamma\pi^+\pi^-\eta$
interactions.
The five-dimensional surface term generates interactions among five
or more Goldstone bosons.

\section{Low-energy phenomenology at $O(p^4)$}
\label{sec:phenomenology}

At lowest order in momenta,
the predictive power of the chiral Lagrangian was
really impressive: with only two low-energy couplings, it was
possible to describe all Green functions associated
with the pseudoscalar-meson interactions.
The symmetry constraints become less powerful at
higher orders. Ten additional constants appear in the
$\cL_4$ Lagrangian, and many more would be needed
at $O(p^6)$.
Higher-order terms in the chiral expansion
are much more sensitive
to the non-trivial aspects of the underlying QCD dynamics.

With $p \lsim M_K \, (M_\pi)$,
we expect $O(p^4)$
corrections to the lowest-order amplitudes at the level
of $p^2/\Lambda_\chi^2 \lsim 20\% \, (2\% )$.
We need to include those corrections if we aim to increase
the accuracy of the ChPT predictions beyond this level.
Although the number of free constants in $\cL_4$ looks
quite big, only a few of them
contribute to a given  observable.
In the absence of external fields, for instance,
the Lagrangian reduces to the first three terms; elastic
$\pi\pi$ and $\pi K$ scatterings are then sensitive to
$L_{1,2,3}$.
The two-derivative couplings $L_{4,5}$ generate mass corrections
to the meson decay constants (and
mass-dependent wave-function renormalizations).
Pseudoscalar masses are affected by the non-derivative
terms $L_{6,7,8}$;
$L_9$ is mainly responsible for the charged-meson electromagnetic radius
and $L_{10}$, finally, only contributes to amplitudes with at least
two external vector or axial-vector fields, like
the radiative semileptonic decay $\pi\to e\nu\gamma$.

Table \ref{tab:Lcouplings},
taken from ref. \cite{ref:BEG92},
summarizes the present
status of the phenomenological determination \cite{ref:GL85,ref:BI88}
of the constants $L_i$.
The quoted numbers correspond to the
renormalized couplings, at a scale $\mu = M_\rho$.
The values of these couplings at any other renormalization scale can be
trivially obtained, through the logarithmic running implied by
Eq. (\ref{eq:renormalization}):
\be
L_i^r(\mu_2) \, = \, L_i^r(\mu_1) \, + \, {\Gamma_i\over (4\pi)^2}
\,\log{\left({\mu_1\over\mu_2}\right)} .
\ee
%

\begin{table}
\caption{Phenomenological values of the
renormalized couplings $L_i^r(M_\rho)$.
The last column shows the source used to extract this information.}
\vspace{0.2cm}
\label{tab:Lcouplings}
\begin{center}
\begin{tabular}{|c|c|c|}
\hline
$i$ & $L_i^r(M_\rho) \times 10^3$ & Source
\\ \hline
1 & $\hphantom{-}0.7\pm0.5$ & $K_{e4}$, $\pi\pi\to\pi\pi$
\\
2 & $\hphantom{-}1.2\pm0.4$ & $K_{e4}$, $\pi\pi\to\pi\pi$
\\
3 & $-3.6\pm1.3$ & $K_{e4}$, $\pi\pi\to\pi\pi$
\\
4 & $-0.3\pm0.5$ &  Zweig rule
\\
5 & $\hphantom{-}1.4\pm0.5$ & $F_K : F_\pi$
\\
6 & $-0.2\pm0.3$ & Zweig rule
\\
7 & $-0.4\pm0.2$ & Gell-Mann--Okubo, $L_5$, $L_8$
\\
8 & $\hphantom{-}0.9\pm0.3$ & $M_{K^0} - M_{K^+}$, $L_5$,
$(m_s - \hat{m}) : (m_d-m_u)$
\\
9 & $\hphantom{-}6.9\pm0.7$ & $\langle r^2\rangle^\pi_{\rm em}$
\\
10 & $-5.5\pm0.7$ & $\pi\to e\nu\gamma$
\\ \hline
\end{tabular}
\end{center}
\end{table}

Comparing the Lagrangians $\cL_2$ and $\cL_4$, one can make an estimate
of the expected size of the couplings $L_i$ in terms of the scale of
SCSB. Taking
$\Lambda_\chi \sim 4 \pi f_\pi \sim 1.2\, \mbox{\rm GeV}$,
one would get
\be
L_i \sim {f_\pi^2/4 \over \Lambda_\chi^2} \sim {1\over 4 (4 \pi)^2}
\sim 2\times 10^{-3} ,
\ee
in reasonable agreement with the phenomenological values quoted in
Table~\ref{tab:Lcouplings}.
This indicates a good convergence of the momentum expansion
below the  resonance region, i.e. $p < M_\rho$.

The chiral Lagrangian allows us to make a good book-keeping of
phenomenological information with a few couplings.
Once these couplings have been fixed,
we can predict many other quantities. Moreover,
the information contained in Table~\ref{tab:Lcouplings}
is very useful to easily test different QCD-inspired models.
Given any particular model aiming to correctly describe QCD at low
energies, we no longer need to make an extensive
phenomenological analysis of all the predictions
of the model, in order to test
its degree of reliability;
we only need
to calculate the predicted
low-energy couplings, and compare them with
the values in Table~\ref{tab:Lcouplings}.
For instance, if one integrates out the heavy scalar of
the linear sigma model described in Section~\ref{sec:sigma},
the resulting Goldstone Lagrangian
only contains the $L_1$ term [see Eq.~(\ref{eq:sigma5})]
at $O(p^4)$;
obviously this is not a satisfactory approximation to  the
physical world\footnote{
A more detailed study of the renormalizable linear sigma model
can be found in ref. \protect\cite{ref:GL84}.
The conclusion is that this model is clearly ruled out by the data.}.

An exhaustive description of the chiral phenomenology
at $O(p^4)$  is beyond the
scope of these lectures.
Instead, I will just present a few examples to
illustrate both the power and limitations of the ChPT techniques.

\subsection{Decay constants}

In the isospin limit ($m_u = m_d = \hat{m}$),
the $O(p^4)$ calculation of the meson-decay constants
gives \cite{ref:GL85}:
\beqn
f_\pi & = & f \left\{ 1 - 2\mu_\pi - \mu_K +
    {4 M_\pi^2\over f^2} L_5^r(\mu)
    + {8 M_K^2 + 4 M_\pi^2 \over f^2} L_4^r(\mu)
    \right\} , \no\\
f_K & = & f \left\{ 1 - {3\over 4}\mu_\pi - {3\over 2}\mu_K
    - {3\over 4}\mu_{\eta_8}
    + {4 M_K^2\over f^2} L_5^r(\mu)
    + {8 M_K^2 + 4 M_\pi^2 \over f^2} L_4^r(\mu)
    \right\} , \quad\\
f_{\eta_8} & = & f \left\{ 1 - 3\mu_K +
    {4 M_{\eta_8}^2\over f^2} L_5^r(\mu)
    + {8 M_K^2 + 4 M_\pi^2 \over f^2} L_4^r(\mu)
    \right\} , \no
\eeqn
where
\be
\mu_P \equiv {M_P^2\over 32 \pi^2 f^2} \,
\log{\left( {M_P^2\over\mu^2}\right)} .
\ee
The result depends on two $O(p^4)$ couplings, $L_4$ and $L_5$.
The $L_4$ term generates a universal shift of all meson-decay
constants,
$\delta f^2 = 16 L_4 B_0 \langle\cM\rangle$,
which can be eliminated taking ratios.
{}From the experimental value \cite{ref:LR84}
\be
{f_K\over f_\pi} = 1.22\pm 0.01 ,
\ee
one can then fix $L_5(\mu)$; this gives the result quoted in
Table~\ref{tab:Lcouplings}.
Moreover, one gets the absolute prediction \cite{ref:GL85}
\be
{f_{\eta_8}\over f_\pi} = 1.3 \pm 0.05 .
\ee
Taking into account isospin violations, one can also predict
\cite{ref:GL85} a tiny
difference between $f_{K^\pm}$ and $f_{K^0}$, proportional to
$m_d-m_u$.

\subsection{Electromagnetic form factors}

At $O(p^2)$ the electromagnetic coupling of the Goldstone bosons
is just the minimal one, obtained through the covariant derivative.
The next-order corrections generate a momentum-dependent
form factor
\be
F^{\phi^\pm}(q^2) = 1 + {1\over 6} \,
\langle r^2 \rangle^{\phi^\pm} \, q^2 + \ldots \, ;
\qquad\qquad
F^{\phi^0}(q^2) =  {1\over 6} \,
\langle r^2 \rangle^{\phi^0} \, q^2 + \ldots
\ee
The meson electromagnetic radius
$\langle r^2 \rangle^\phi$
gets local contributions from the $L_9$ term,
plus logarithmic loop corrections \cite{ref:GL85}:
\beqn\label{eq:radius}
\langle r^2 \rangle^{\pi^\pm} & = & {12 L^r_9(\mu)\over f^2}
    - {1\over 32 \pi^2 f^2} \left\{
   2 \log{\left({M_\pi^2\over\mu^2}\right)}
    + \log{\left({M_K^2\over\mu^2}\right)} + 3 \right\} ,
\no\\
\langle r^2 \rangle^{K^0} & = & - {1\over 16 \pi^2 f^2}
\,\log{\left({M_K\over M_\pi}\right) } ,
\\
\langle r^2 \rangle^{K^\pm} & = & \langle r^2 \rangle^{\pi^\pm}
   + \langle r^2 \rangle^{K^0} . \no
\eeqn

Since neutral bosons do not couple to the photon at tree level,
$\langle r^2 \rangle^{K^0}$
only gets a loop contribution, which is moreover finite
(there cannot be any divergence because there
exists no counter-term to renormalize it!).
The predicted value,
$\langle r^2 \rangle^{K^0} = -0.04\pm 0.03 \, \mbox{\rm fm}^2$, is in
perfect agreement with the experimental determination \cite{ref:MO78}
$\langle r^2 \rangle^{K^0} = -0.054\pm 0.026 \, \mbox{\rm fm}^2$.

The measured electromagnetic pion radius \cite{ref:AM86},
$\langle r^2 \rangle^{\pi^\pm} = 0.439\pm 0.008 \, \mbox{\rm fm}^2$,
is used as input to estimate the coupling $L_9$.
This observable provides  a good example of the importance of
higher-order local terms in the chiral expansion.
If one tries to ignore the $L_9$ contribution, using instead some
``physical'' cut-off $p_{\mbox{\rms max}}$ to regularize the  loops,
one needs \cite{ref:LE89}
   $p_{\mbox{\rms max}}\sim 60 \,\mbox{\rm GeV}$,
in order to reproduce the experimental value; this is clearly
nonsense.
The pion charge radius is dominated by the $L^r_9(\mu)$
contribution, for any reasonnable value of $\mu$.

The measured $K^+$ charge radius \cite{ref:DA82},
$\langle r^2 \rangle^{K^\pm} = 0.28\pm 0.07 \, \mbox{\rm fm}^2$,
has a larger experimental uncertainty.
Within present errors, it is in agreement with the parameter-free
relation in Eq.~(\ref{eq:radius}).

\subsection{$K_{l3}$ decays}

The semileptonic decays $K^+\to\pi^0 l^+ \nu_l$ and
$K^0\to\pi^- l^+ \nu_l$ are governed by the corresponding
hadronic matrix
element of the vector current [$t\equiv (P_K-P_\pi)^2$],
\be
\langle \pi| \bar s\gamma^\mu u |K\rangle = C_{K\pi} \,\left[
\left( P_K + P_\pi\right)^\mu \, f_+^{K\pi}(t) \, + \,
\left( P_K - P_\pi\right)^\mu \, f_-^{K\pi}(t) \right] ,
\ee
where $C_{K^+\pi^0} = 1/\sqrt{2}$, $C_{K^0\pi^-} = 1$.
At lowest order, the two form factors reduce to trivial constants:
$f_+^{K\pi}(t) = 1$ and $f_-^{K\pi}(t) = 0$.
There is however a sizeable correction to $f_+^{K^+\pi^0}(t)$,
due to $\pi^0\eta$ mixing, which is
proportional to $(m_d-m_u)$,
\be
f_+^{K^+\pi^0}(0) \, = \,
1 + {3\over 4} \, {m_d-m_u\over m_s - \hat{m}}
\, = \, 1.017 .
\ee
This number should be compared with the experimental ratio
\be\label{eq:expratio}
{f_+^{K^+\pi^0}(0)\over f_+^{K^0\pi^-}(0)} \, = \,
1.028 \pm 0.010 .
\ee
The $O(p^4)$ corrections to $f_+^{K\pi}(0)$ can be expressed in
a parameter-free manner in terms of the physical meson masses
\cite{ref:GL85}.
Including those contributions,
one gets the more precise values
\be
 f_+^{K^0\pi^-}(0) = 0.977 , \qquad \qquad
{f_+^{K^+\pi^0}(0)\over f_+^{K^0\pi^-}(0)} = 1.022 ,
\ee
which are in perfect agreement with the experimental result
(\ref{eq:expratio}).
The accurate ChPT calculation of these quantities allows us to
extract \cite{ref:LR84} the most precise determination
of the Kobayashi--Maskawa matrix element $V_{us}$:
\be
|V_{us}| = 0.2196 \pm 0.0023 .
\ee

At $O(p^4)$, the form factors get momentum-dependent contributions.
Since $L_9$ is
the only unknown chiral coupling occurring in $f_+^{K\pi}(t)$ at this
order, the slope
$\lambda_+$
of this form factor can be fully predicted.
Alternatively, we can use the measured slope \cite{ref:PDG92},
\be\label{eq:slope}
\lambda_+ \equiv {1\over 6 }\,\langle r^2\rangle^{K\pi}\, M_\pi^2
= 0.0300\pm 0.0016 ,
\ee
as an input to get an independent determination of $L_9$.
The value (\ref{eq:slope}) corresponds \cite{ref:GL85} to
$L_9^r(M_\rho) = (6.6\pm0.4)\times 10^{-3}$, in excellent agreement
with the determination from the pion-charge radius, quoted in
Table \ref{tab:Lcouplings}.

Instead of $f_-^{K\pi}(t)$, it is usual to parametrize the
experimental results in terms of the so-called
scalar form factor
\be
f_0^{K\pi}(t) = f_+^{K\pi}(t) +{t\over M_K^2 - M_\pi^2} f_-^{K\pi}(t) .
\ee
The slope of this form factor is determined by the constant $L_5$,
which in turn is fixed by $f_K/f_\pi$.
One gets the result \cite{ref:GL85}:
\be\label{eq:slope2}
\lambda_0 \equiv {1\over 6 }\,\langle r^2\rangle^{K\pi}_S\, M_\pi^2
= 0.017\pm 0.004 .
\ee
The experimental situation concerning the
value of this slope is far from clear;
while an older high-statistics measurement \cite{ref:DO74},
$\lambda_0 = 0.019\pm 0.004$, confirmed the theoretical expectations,
more recent experiments find higher values, which disagree with this
result. Reference
\cite{ref:CHO80}, for instance, report
$\lambda_0 = 0.046\pm 0.006$,
which differs from (\ref{eq:slope2}) by more than 4 standard
deviations.
The Particle Data Group \cite{ref:PDG92} quote a world average
$\lambda_0 = 0.025\pm 0.006$.

\subsection{Meson masses}

The relations (\ref{eq:masses}) get modified at $O(p^4)$.
The additional contributions depend on the low-energy constants
$L_4$, $L_5$, $L_6$, $L_7$ and $L_8$.
It is possible, however, to obtain one relation between the
quark and meson masses, which does not contain any of the $O(p^4)$
couplings.
The dimensionless ratios
\be\label{eq:q1q2_def}
Q_1 \equiv {M_K^2 \over M_\pi^2} , \qquad\qquad
Q_2 \equiv {(M_{K^0}^2 - M_{K^+}^2) - (M_{\pi^0}^2 - M_{\pi^+}^2)
    \over M_K^2 - M_{\pi}^2} ,
\ee
get  the same $O(p^4)$ correction \cite{ref:GL85}:
\be\label{eq:q1q2}
Q_1 = {m_s + \hat{m} \over 2 \hat{m}} \, \{ 1 + \Delta_M\} ,
\qquad\qquad
Q_2 = {m_d - m_u \over m_s - \hat{m}} \, \{ 1 + \Delta_M\} ,
\ee
where
\be
\Delta_M = - \mu_\pi + \mu_{\eta_8} + {8\over f^2}\,
(M_K^2 - M_\pi^2)\, \left[ 2 L_8^r(\mu) - L_5^r(\mu)\right] .
\ee
Therefore, at this order, the ratio $Q_1/Q_2$ is just given
by the corresponding
ratio of quark masses,
\be\label{eq:Q2}
Q^2 \equiv {Q_1\over Q_2} =
{m_s^2 - \hat{m}^2 \over m_d^2 - m_u^2} .
\ee
To a good approximation, Eq.~(\ref{eq:Q2})
can be written as an ellipse,
\be
\left({m_u\over m_d}\right)^2 + {1\over Q^2}\,
\left({m_s\over m_d}\right)^2 = 1 ,
\ee
which constrains the quark-mass ratios.
The observed values of the meson masses give $Q = 24$.

Obviously, the quark-mass ratios (\ref{eq:Weinbergratios}),
obtained at $O(p^2)$, satisfy this elliptic constraint.
At $O(p^4)$, however, it is not possible to make a separate
determination of $m_u/m_d$ and $m_s/m_d$ without having additional
information on some of the $L_i$ couplings.

A useful quantity is the deviation of the Gell-Mann--Okubo relation,
\be
\Delta_{\rm GMO} \equiv {4 M_K^2 - 3 M_{\eta_8}^2 - M_\pi^2
    \over M_{\eta_8}^2 - M_\pi^2} .
\ee
Neglecting the mass difference $m_d-m_u$, one gets \cite{ref:GL85}
\beqn\label{eq:dGMO}
\Delta_{\rm GMO} & = &
 {-2 \,(4 M_K^2 \mu_K - 3 M_{\eta_8}^2 \mu_{\eta_8} - M_\pi^2 \mu_\pi)
    \over M_{\eta_8}^2 - M_\pi^2}
\no\\ & &
   -{6\over f^2} \, (M_{\eta_8}^2 - M_\pi^2)
   \, \left[ 12 L_7^r(\mu) + 6 L_8^r(\mu) - L_5^r(\mu)\right] .
\eeqn
Experimentally, correcting the masses for electromagnetic effects,
$\Delta_{\rm GMO} = 0.21$. Since $L_5$ is already known, this allows
the combination $2 L_7 + L_8$ to be fixed .

In order to determine the individual quark-mass ratios
from Eqs.~(\ref{eq:q1q2}), we
would need to fix
the constant $L_8$. However, there is no way to find an observable
that isolates this coupling.
The reason is an accidental symmetry of the Lagrangian $\cL_2 + \cL_4$.
The chiral Lagrangian remains
invariant under
the following simultaneous change \cite{ref:KM90}
of the quark-mass matrix
and some of the chiral couplings:
\beqn\label{eq:kmsymmetry}
\cM' & = & \alpha \cM + \beta (\cM^\dagger)^{-1} \, \det\cM ,
\qquad B_0' \, = \, B_0 / \alpha ,
\\
L'_6 & = & L_6 - \zeta , \qquad
L'_7 \, = \, L_7 - \zeta , \qquad
L'_8 \, = \, L_8 + 2 \zeta , \quad \no
\eeqn
where $\alpha$ and $\beta$ are arbitrary constants, and
$\zeta = \beta f^2 / (32\alpha B_0)$.
The only information on the quark-mass matrix $\cM$ that we used
to construct the effective Lagrangian was that it transforms as
$\cM\to g_R \cM g_L^\dagger$.
The matrix $\cM'$ transforms in the same manner;
therefore, symmetry alone does not allow us to distinguish between
$\cM$ and $\cM'$.
Since only the product $B_0 \cM$ appears in the Lagrangian,
$\alpha$ merely changes the value of the constant $B_0$.
The term proportional to $\beta$ is a correction of $O(\cM^2)$;
when inserted in $\cL_2$, it generates a contribution to
$\cL_4$, which is reabsorbed by the redefinition
of the $O(p^4)$ couplings.
All chiral predictions will be invariant under the transformation
(\ref{eq:kmsymmetry}); therefore it is not possible to
separately determine the values of the quark masses and the
constants $B_0$, $L_6$, $L_7$ and $L_8$.
We can only fix those combinations of chiral couplings and masses
that remain invariant under (\ref{eq:kmsymmetry}).

Notice that  (\ref{eq:kmsymmetry})
is certainly not a symmetry of the underlying
QCD Lagrangian.
The accidental symmetry arises in the effective theory
because we are not making use of the explicit form of the
QCD Lagrangian; only its symmetry properties under chiral rotations
have been taken into account.
Therefore, we can resolve the ambiguity by obtaining
one additional information from outside the pseudoscalar-meson
chiral Lagrangian framework.
For instance, by analysing the
isospin breaking in the baryon mass spectrum and the $\rho$-$\omega$
mixing \cite{ref:GL82}, it is possible to fix the ratio
$(m_s - \hat{m})/ (m_d - m_u) = 43.7\pm 2.7$.
Inserting this number in Eq.~(\ref{eq:Q2}), one gets  \cite{ref:GL85}
\be
{m_s \over \hat{m}} = 25.7 \pm 2.6 .
\ee
Moreover, one can now determine  $L_8$ from
Eqs. (\ref{eq:q1q2}), and  therefore fix $L_7$ with Eq.~(\ref{eq:dGMO});
one then gets the values quoted in Table \ref{tab:Lcouplings}.
Other ways of resolving the ambiguity, by using different additional
inputs \cite{ref:LE90,ref:DW92}, lead to similar estimates of the
quark-mass ratios and the low-energy couplings.

\section{Information encoded in the chiral couplings}
\label{sec:couplings}

The effective theory takes
explicitly into account the poles and
cuts generated by the Goldstone bosons.
Given the non-trivial analytic structure associated with
those physical
intermediate states,
 the full amplitudes are
reconstructed up to a subtraction polynomial.
Obviously, the subtraction constants $L_i$ contain all the information
on the heavy degrees of freedom, which do not appear in the low-energy
Lagrangian.

It seems rather natural to expect that the lowest-mass resonances,
such as $\rho$ mesons, should have an important impact on the physics
of the pseudoscalar bosons.
In particular, the low-energy singularities due to the exchange of
those resonances should generate sizeable contributions to the
chiral couplings.
This can be easily understood, making a Taylor expansion of the
$\rho$ propagator:
\be
{1\over p^2 - M_\rho^2} \, = \, {-1\over M_\rho^2} \,
\left\{ 1 + {p^2\over M_\rho^2} + \ldots \right\},
\qquad\qquad (p^2 < M_\rho^2) .
\ee
Below the $\rho$-mass scale, the singularity associated with the pole
of the resonance propagator is replaced by the corresponding
momentum expansion.
The exchange of virtual $\rho$ mesons should result in derivative
Goldstone couplings proportional to powers of $1/M_\rho^2$.

It is well known, for instance, that the electromagnetic form factor
of the charged pion is well reproduced by the vector-meson dominance (VMD)
formula
\be\label{eq:vmd}
F^{\pi^\pm}(t) \approx {M_\rho^2 \over M_\rho^2 -t } ,
\ee
i.e.
$\langle r^2\rangle^{\pi^\pm} \approx 6/M_\rho^2 = 0.4 \,\hbox{\rm fm}^2$,
to be compared with the measured value
$\langle r^2 \rangle^{\pi^\pm} = 0.439\pm 0.008 \, \mbox{\rm fm}^2$.

Writing a chiral-invariant $\rho\pi\pi$ interaction, with coupling
$G_V$, one can  compute the effect of a $\rho$-exchange
diagram at low energies; the leading contribution
\cite{ref:GL84,ref:EGPR89}
 is a $\pi^4$
local interaction, with a coupling constant proportional to
$G_V^2/M_\rho^2$.
Since $G_V$ can be directly measured from the $\rho\to 2\pi$ decay
width, $|G_V| = 69$ MeV,
the size of this contribution is fully predicted.
Similarly, one can write
 a chiral invariant $\rho^0\gamma$ interaction,
with coupling $F_V$; this coupling can be extracted from the
$\rho^0\to e^+e^-$ decay width,
$|F_V| = 154$ MeV.
The exchange of a $\rho$ meson between the $G_V$ and $F_V$ vertices,
generates a contribution to the electromagnetic form factor of the
charged pion \cite{ref:EGPR89}:
\be
\langle r^2\rangle^{\pi^\pm} = {6 F_V G_V \over f^2 M_\rho^2} .
\ee
{}From the success of the na\"{\i}ve  VMD formula (\ref{eq:vmd}),
one could expect $F_V G_V / f^2 \approx 1$, which is indeed
approximately satisfied (one obtains 1.2 with the measured
$F_V$ and $G_V$ values).

A systematic analysis of the
role of resonances in the ChPT Lagrangian
has been performed\footnote{
Related work can be found in ref. \protect\cite{ref:DRV89}.}
in ref. \cite{ref:EGPR89}.
One writes first a general chiral-invariant Lagrangian
$\cL(U,V,A,S,P)$,
describing
the couplings between meson resonances of the type $V$, $A$, $S$, $P$
and the Goldstone bosons, at lowest-order in derivatives.
The coupling constants of this Lagrangian are
phenomenologically extracted from physics at the resonance-mass scale.
One has then an effective chiral theory defined in the
intermediate-energy region.
Formally, the generating functional (\ref{eq:generatingfunctional})
is given in this theory by the path-integral formula
\be
\exp{\{i Z\}} \, = \,
\int \, {\cal D}U(\phi)\, \cD V \,\cD A \,\cD S \,\cD P
\, \exp{\left\{ i \int d^4x \,\cL(U,V,A,S,P) \right\}} .
\ee

The integration of the resonance fields  results
in a low-energy theory with  only Goldstones, i.e.
the usual ChPT Lagrangian.
At lowest-order this integration can be explicitly performed,
expanding around the classical solution for the resonance fields.
The  resulting $L_i$ couplings
\cite{ref:EGPR89}
are summarized in Table \ref{tab:vmd},
which compares the different resonance-exchange contributions
with the phenomenologically determined values of $L_i^r(M_\rho)$.
For vector and axial-vector mesons only the $SU(3)$ octets
contribute, whereas
both octets and singlets are relevant
in the case
of scalar and pseudoscalar resonances.

\begin{table}
\caption{$V$, $A$, $S$, $S_1$ and $\eta_1$ contributions to the
 coupling constants $L_i^r$ in units of $10^{-3}$.
 The last column shows the results obtained with the relations
 in Eq.~(\protect\ref{eq:vmdpred}).}
\vspace{0.2cm}
\label{tab:vmd}
\begin{center}
\begin{tabular}{|c|c||ccccc|c||c|}
\hline
i & $L_i^r(M_\rho)$ & $\hphantom{.0}V$ & $A\,$ & $\,S$ &
      $S_1$ & $\eta_1$ & Total & Total$^{c)}$
\\ \hline
1 & $\hphantom{-}0.7\pm0.5$ &
      $\hphantom{-1}0.6$ & $0\hphantom{.0}$ & $-0.2$ &
      $0.2^{b)}$ & $0$ & $\hphantom{-}0.6$ & $\hphantom{-}0.9$
\\
2 & $\hphantom{-}1.2\pm0.4$ &
      $\hphantom{-1}1.2$ & $0\hphantom{.0}$ & $0$ &
      $0\hphantom{.2^{b)}}$ & $0$ & $\hphantom{-}1.2$
      & $\hphantom{-}1.8$
\\
3 & $-3.6\pm1.3$ &
      $\,-3.6$ & $0\hphantom{.0}$ & $\hphantom{-}0.6$ &
      $0\hphantom{.2^{b)}}$ & $0$ & $-3.0$ & $-4.9$
\\
4 & $-0.3\pm0.5$ &
      $\hphantom{-1}0\hphantom{.0}$ & $0\hphantom{.0}$ & $-0.5$ &
      $0.5^{b)}$ & $0$ & $\hphantom{-}0.0$ & $\hphantom{-}0.0$
\\
5 & $\hphantom{-}1.4\pm0.5$ &
      $\hphantom{-1}0\hphantom{.0}$ & $0\hphantom{.0}$ &
      $\hphantom{-1}1.4^{a)}$ &
      $0\hphantom{.2^{b)}}$ & $0$ & $\hphantom{-}1.4$
      & $\hphantom{-}1.4$
\\
6 & $-0.2\pm0.3$ &
      $\hphantom{-1}0\hphantom{.0}$ & $0\hphantom{.0}$ & $-0.3$ &
      $0.3^{b)}$ & $0$ & $\hphantom{-}0.0$ & $\hphantom{-}0.0$
\\
7 & $-0.4\pm0.2$ &
      $\hphantom{-1}0\hphantom{.0}$ & $0\hphantom{.0}$ & $0$ &
      $0\hphantom{.2^{b)}}$ & $-0.3$ & $-0.3$ & $-0.3$
\\
8 & $\hphantom{-}0.9\pm0.3$ &
      $\hphantom{-1}0\hphantom{.0}$ & $0\hphantom{.0}$ &
      $\hphantom{-1}0.9^{a)}$ &
      $0\hphantom{.2^{b)}}$ & $0$ & $\hphantom{-}0.9$
      & $\hphantom{-}0.9$
\\
9 & $\hphantom{-}6.9\pm0.7$ &
      $\hphantom{-11}6.9^{a)}$ & $0\hphantom{.0}$ & $0$ &
      $0\hphantom{.2^{b)}}$ & $0$ & $\hphantom{-}6.9$
      & $\hphantom{-}7.3$
\\
10 & $-5.5\pm0.7$ &
      $-10.0$ & $4.0$ & $0$ &
      $0\hphantom{.2^{b)}}$ & $0$ & $-6.0$ & $-5.5$
\\ \hline
\end{tabular}

\hbox{$\hphantom{B}$}
\hbox{$\qquad\qquad\quad$ $^{a)}$ Input. $\qquad$
$^{b)}$ Large-$N_C$ estimate. $\qquad$
$^{c)}$ With (\protect\ref{eq:vmdpred})}
\end{center}
\end{table}

At lowest order, the most general interaction of the
$V$ octet to the Goldstone bosons contains two terms,
corresponding to the couplings $G_V$ and $F_V$
described before.
Due to the different parity, only one term with coupling
$F_A$ is present for the axial octet $A$. While
$V$ exchange generates contributions
to
$L_1$, $L_2$, $L_3$, $L_9$ and $L_{10}$,
$A$ exchange only contributes
to $L_{10}$ \cite{ref:EGPR89}:
\beqn
&& L_1^V = {G_V^2\over 8 M_V^2}, \qquad L_2^V = 2 L_1^V ,
\qquad L_3^V = -6 L_1^V , \\
&& L_9^V = {F_V G_V\over 2 M_V^2}, \qquad
L_{10}^{V+A} = - {F_V^2\over 4 M_V^2} + {F_A^2\over 4 M_A^2} . \no
\eeqn
To obtain the numbers in Table~\ref{tab:vmd},
the value of $L_9^r(M_\rho)$ has been fitted to determine
$|G_V| = 53 \,\mbox{\rm MeV}$;
nevertheless, the qualitative conclusion would be the same
with the $\rho\to 2 \pi$ determination mentioned before.
The axial parameters have been fixed using the old Weinberg
sum rules \cite{ref:WE67}:
$F_A^2 = F_V^2 - f_\pi^2 = (123 \, \mbox{\rm MeV})^2$
and $M_A^2 = M_V^2 F_V^2/ F_A^2 = (968 \,\mbox{\rm MeV})^2$.
The results shown in the table clearly establish a chiral
version of vector (and axial-vector) meson dominance:
whenever they can contribute at all, $V$ and $A$ exchange
seem to completely dominate the relevant coupling constants.

There are different phenomenologically successful models
in the literature
for $V$ and $A$ resonances
(tensor-field description
\cite{ref:GL84,ref:EGPR89},
massive Yang--Mills \cite{ref:ME88},
hidden gauge formulations \cite{ref:BKY88}, etc.).
It can be shown \cite{ref:EGLPR89}   that all models are equivalent (i.e.
give the same contributions to the $L_i$),
provided they incorporate the appropriate
QCD constraints at high energies.
Moreover, with additional QCD-inspired assumptions of
high-energy behaviour, such as an unsubtracted dispersion relation
for the pion electromagnetic form factor,
all $V$ and $A$ couplings can be expressed in terms of
$f$ and $M_V$ only \cite{ref:EGLPR89}:
\be\label{eq:f_relations}
F_V = \sqrt{2} f_\pi, \qquad G_V= f_\pi/\sqrt{2}, \qquad
F_A=f_\pi, \qquad M_A=\sqrt{2} M_V.
\ee
In that case, one has
\be\label{eq:vmdpred}
L_1^V = L_2^V/2 = - L_3^V/6 = L_9^V/8 = -L_{10}^{V+A}/6
= f_\pi^2/(16 M_V^2).
\ee
The last column in Table~\ref{tab:vmd} shows the
predicted numerical values of the $L_i$ couplings, using
the relations (\ref{eq:vmdpred}).

The analysis of scalar exchange is very similar \cite{ref:EGPR89}.
Since the experimental information is
quite scarce in the scalar sector,
one needs to assume that
the couplings $L_5$ and $L_8$ are due exclusively
to scalar-octet exchange, to determine the scalar-octet couplings.
The scalar-octet contributions to the other $L_i$ ($i=1,3,4,6$)
are then fixed. Moreover, one can then predict
$\Gamma(a_0\to\eta\pi)$, in good agreement with experiment.
The scalar-singlet-exchange contributions can be expressed in
terms of the octet parameters using large-$N_C$ arguments.
For $N_C=\infty$, octet- and singlet-scalar exchange cancel
in $L_1$, $L_4$ and $L_6$.
Although the results in Table~\ref{tab:vmd}
cannot be considered as a proof for scalar dominance,
they provide at least a convincing demonstration of its consistency.

Neglecting the higher-mass
$0^-$ resonances,
the only  remaining meson-exchange is the one associated with
the $\eta_1$, which generates a sizeable contribution to $L_7$
\cite{ref:GL85,ref:EGPR89}.
The magnitude of this contribution can be calculated from
the quark-mass expansion of $M_\eta^2$.
The result for $L_7$ is in close agreement with its phenomenological
value.

The combined resonance contributions appear to saturate the $L_i^r$
almost entirely \cite{ref:EGPR89}.
Within the uncertainties of the approach,
there is no need for invoking any additional contributions.
Although the comparison has been made for $\mu=M_\rho$,
a similar conclusion would apply for any value of $\mu$
in the low-lying resonance region between 0.5 and 1 GeV.

All chiral couplings are in principle calculable from QCD.
They are functions of $\Lambda_{QCD}$ and the heavy-quark
masses $m_c$, $m_b$, $m_t$. Unfortunately, we are not able
at present to make such a first-principle computation.
Although the integral over the quark fields in
Eq.~(\ref{eq:generatingfunctional})
can be done explicitly, we do not know how to
perform analytically the remaining integration over the
gluon fields.
A perturbative evaluation of the gluonic contribution would
obviously fail in reproducing the correct dynamics of SCSB.
A possible way out is to parametrize phenomenologically the
SCSB and make a weak gluon-field expansion around the
resulting physical vacuum.

The simplest parametrization \cite{ref:ERT90}
is obtained by adding to the QCD Lagrangian the term
\be\label{eq:ERTmodel}
\Delta\cL_{\rm QCD} = - M_Q \left( \bar q_R U q_L +
    \bar q_L U^\dagger q_R \right) ,
\ee
which serves to introduce the $U$ field, and a mass parameter
$M_Q$, which regulates the infra-red behaviour of the low-energy
effective action. In the presence of this term
the operator $\bar q q$ acquires a vacuum expectation value;
therefore, (\ref{eq:ERTmodel}) is an effective way to
generate the order parameter due to SCSB.
Making a chiral rotation of the quark fields,
$Q_L = \xi q_L$, $Q_R = \xi^\dagger q_R$, with $\xi$ chosen such that
$U=\xi^2$,
the interaction (\ref{eq:ERTmodel}) reduces to a
mass-term for the ``dressed'' quarks $Q$; the parameter
$M_Q$ can then be interpreted as a
``constituent-quark mass''.

The derivation of the low-energy effective chiral Lagrangian
within this framework has been extensively discussed in
ref. \cite{ref:ERT90}.
In the chiral and large-$N_C$ limits,
and including the leading gluonic
contributions, one gets:
\beqn\label{eq:ERTresult}
&& 8 L_1 = 4 L_2 = L_9 = {N_C\over 48\pi^2}
  \left[ 1 + O\left(1/M_Q^6\right)\right] , \\
&& L_3 = L_{10} = -{N_C\over 96\pi^2}
   \left[ 1 + {\pi^2\over 5 N_C}
   {\langle{\alpha_s\over\pi}GG\rangle\over M_Q^4} +
O\left(1/M_Q^6\right)\right] . \qquad\no
\eeqn
Due to dimensional reasons, the leading contributions
to the $O(p^4)$ couplings only depend on
$N_C$ and geometrical factors.
It is remarkable  that $L_1$, $L_2$ and $L_9$
do not get any gluonic correction at this order; this
result is independent of the way SCSB has been
parametrized
($M_Q$ can be taken to be infinite).
Table~\ref{tab:ERT90} compares the predictions obtained with
only the leading term in Eqs.~(\ref{eq:ERTresult})
(i.e. neglecting the gluonic correction) with the
phenomenological determination of the
$L_i$ couplings.
The numerical agreement is quite impressive;
both the order of magnitude and the sign are correctly
reproduced (notice that this is just a free-quark result!).
Moreover, the gluonic corrections shift the values of
$L_3$ and $L_{10}$ in the right direction, making them
more negative.

\begin{table}
\caption{Leading-order ($\alpha_s=0$) predictions for the
$L_i$'s, within the QCD-inspired model (\protect\ref{eq:ERTmodel}).
The phenomenological values are shown in the second row for comparison.
All numbers are given in units of $10^{-3}$.}
\vspace{0.2cm}
\label{tab:ERT90}
\begin{center}
\begin{tabular}{|c|ccccc|}
\hline
& $L_1$ & $L_2$ & $L_3$ & $L_9$ & $L_{10}$
\\ \hline
$L_i^{th}(\alpha_s=0)$ & 0.79 & 1.58 & $-3.17$ & 6.33 & $-3.17$
\\
$L_i^r(M_\rho)$ & $0.7\pm0.5$ & $1.2\pm0.4$ & $-3.6\pm1.3$ &
$6.9\pm0.7$ & $-5.5\pm0.7$
\\ \hline
\end{tabular}
\end{center}
\end{table}

The results (\ref{eq:ERTresult})
obey almost all the short-distance relations (\ref{eq:vmdpred}).
Comparing the predictions for $L_{1,2,9}$ in the VMD approach
of Eq.~(\ref{eq:vmdpred}) with the QCD-inspired
ones in Eq.~(\ref{eq:ERTresult}),
one gets a quite good estimate of the $\rho$ mass:
\be
M_V = 2\sqrt{2}\pi f = 830 \, \mbox{\rm MeV} .
\ee

Is it quite easy to prove that the interaction (\ref{eq:ERTmodel})
is equivalent to the mean-field approximation of the
Nambu--Jona--Lasinio (NJL) model,
where SCSB is triggered by four-quark operators.
It has been conjectured recently \cite{ref:BBR93}
that integrating out the quark and gluon fields of QCD,
down to some intermediate scale
$\Lambda_\chi$, gives rise to an extended NJL
Lagrangian.
By introducing collective fields (to be identified later with
the Goldstone fields and
$S$, $V$, $A$ resonances) the model can be transformed into a
Lagrangian bilinear in the quark fields, which can therefore
be integrated out.
One then gets an effective Lagrangian,
describing the couplings of the pseudoscalar bosons to
vector, axial-vector and scalar resonances.
Extending the analysis beyond the mean-field approximation,
ref. \cite{ref:BBR93} obtains predictions for
20 measurable quantities, including the $L_i$'s, in terms of only
4 parameters. The quality of the fits is quite impressive.
Since the model contains all resonances that are known to saturate
the $L_i$ couplings, it is not surprising that one gets an
improvement of the
mean-field-approximation results, specially for
the constants $L_5$ and $L_8$, which are sensitive to
scalar exchange.
What is more important, this analysis clarifies a
potential problem of double-counting:
in certain limits the model approches either the pure
quark-loop predictions of Eqs.~(\ref{eq:ERTresult}) or
the VMD results (\ref{eq:vmdpred}),
but in general it interpolates between these two cases.

\section{$\Delta S=1$ non-leptonic weak interactions}
\label{sec:weak}

The Standard Model predicts strangeness-changing transitions with
$\Delta S=1$ via $W$-exchange between two weak charged currents.
At low energies ($E<<M_W$), the heavy fields $W$, $Z$, $t$, $b$, $c$
can be integrated out. Using standard operator-product-expansion
techniques, the $\Delta S=1$ weak interactions are described by an
effective Hamiltonian \cite{ref:GW79}
\be\label{eq:ds_hamiltonian}
{\cal H}^{\Delta S = 1}_{\mbox{\rms eff}}
 \, = \, {G_F \over \sqrt{2}} V_{ud}^{\hphantom{*}} V_{us}^* \,
\sum_i C_i(\mu) \, Q_i \, + \, \mbox{h.c.} \, ,
\ee
%
which is a sum of local
four-quark operators, constructed with the light ($u, d, s$) quark fields
only,
\beqn\label{eq:four_quark_operators}
&{\displaystyle
Q_1  \,\equiv \, 4 \, (\bar s_L \gamma^\mu d_L)
\, (\bar u_L \gamma_\mu u_L) ,} \quad \qquad\,
&Q_2 \, \equiv \, 4 \, (\bar s_L \gamma^\mu u_L)
\, (\bar u_L \gamma_\mu d_L) ,
\no\\
&{\displaystyle
Q_3  \,\equiv \, 4 \, (\bar s_L \gamma^\mu d_L) \, \sum_{q=u,d,s}
(\bar q_L \gamma_\mu q_L) ,}\quad
&Q_4 \, \equiv \, 4 \, \sum_{q=u,d,s} (\bar s_L \gamma^\mu q_L) \,
(\bar q_L \gamma_\mu d_L) ,\quad\qquad
\\
&{\displaystyle
Q_5 \,\equiv\, 4 \, (\bar s_L \gamma^\mu d_L) \sum_{q=u,d,s} \,
(\bar q_R \gamma_\mu q_R) ,} \quad
&Q_6 \,\equiv \, -8 \, \sum_{q=u,d,s} (\bar s_L q_R)
\, (\bar q_R d_L) ,
\no
\eeqn
modulated by Wilson coefficients $C_i(\mu)$, which are functions of the heavy
$W$, $t$, $b$, $c$ masses and an overall renormalization scale $\mu$.
Only five of these operators are independent, since
$Q_4 = - Q_1 + Q_2 + Q_3$.
{}From the point of view of chiral $SU(3)_L \otimes SU(3)_R$ and isospin
quantum numbers, $Q_- \equiv Q_2 - Q_1$ and $Q_i \; (i=3,4,5,6)$ transform as
($8_L,1_R$) and induce $|\Delta I| = 1/2$ transitions, while
$Q_1 + 2/3 Q_2 - 1/3 Q_3$ transforms like ($27_L,1_R$) and induces both
$|\Delta I| = 1/2$ and $|\Delta I| = 3/2$ transitions.

In the absence of strong interactions, $C_2(\mu) = 1$ and all other Wilson
coefficients vanish. The Standard Electroweak Model then gives rise to
$|\Delta I| = 1/2$ and $|\Delta I| = 3/2$ amplitudes of nearly equal size,
while experimentally the ratio between the two
amplitudes is a factor of 20.
To solve this large discrepancy, QCD effects should be enormous.
The leading $\alpha_s$ corrections indeed give, for $\mu$-values around 1 GeV,
an enhancement by a factor of 2 to 3 of the $Q_-$ Wilson coefficient
with respect to the $Q_+ \equiv Q_2 + Q_1$ one.
Moreover, the gluonic exchanges generate the additional
$|\Delta I| = 1/2$ operators $Q_i \, (i=3,4,5,6)$.
Nevertheless, this by itself is not enough to explain the experimentally
observed rates, without simultaneously appealing to a further enhancement in
the hadronic matrix elements of at least some of the isospin $1/2$ four-quark
operators.
The computation of hadronic matrix elements
at the $K$-mass scale is however a very difficult
non-perturbative problem.

The effect of $\Delta S=1$ non-leptonic weak interactions can be
incorporated in the low-energy chiral theory, as a perturbation
to the strong effective Lagrangian $\cL_{\mbox{\rms eff}}(U)$.
At lowest order in the number of derivatives,
the most general effective bosonic Lagrangian, with the
same $SU(3)_L\otimes SU(3)_R$ transformation properties
as the four-quark Hamiltonian in
Eqs. (\ref{eq:ds_hamiltonian}) and (\ref{eq:four_quark_operators}),
contains two terms\footnote{
One can build an additional octet term with the external $\chi$ field,
$\langle\lambda\left(U^\dagger\chi + \chi^\dagger U\right)\rangle$;
however, this term does not contribute to on-shell amplitudes.}:
\be\label{eq:L2w}
\cL_2^{\Delta S=1} =
-{G_F\over\sqrt{2}} V_{ud}^{\hphantom{*}} V_{us}^*\,\left\{
g_8  \langle \lambda L_\mu L^\mu\rangle +
g_{27} \left( L_{\mu 23} L^\mu_{11} + \frac{2}{3} L_{\mu 21}
L^\mu_{13}\right) + \mbox{\rm h.c.} \right\} ,
\ee
where
\be
\lambda =  (\lambda_6 - i \lambda_7) /2 , \qquad\qquad
L_\mu = i f^2 U^\dg D_\mu U .
\ee
The chiral couplings $g_8$ and $g_{27}$ measure the strength
of the two parts in the effective Hamiltonian (\ref{eq:ds_hamiltonian})
transforming as $(8_L,1_R)$ and $(27_L,1_R)$, respectively,
under chiral rotations.
Their values can be extracted from $K\to 2\pi$ decays
\cite{ref:PGR86}:
\be\label{eq:g8g27}
|g_8| \approx 5.1 , \qquad\qquad
g_{27} / g_8 \approx 1/18 .
\ee
The huge difference between these two couplings shows the
enhancement of the octet $|\Delta I|=1/2$ transitions.

Using the effective Lagrangian (\ref{eq:L2w}),
the calculation of hadronic weak transitions becomes a straightforward
perturbative problem. The highly non-trivial QCD dynamics
has been parametrized in terms of the two chiral couplings.
Of course, the interesting problem that remains to be solved is
to compute $g_8$ and $g_{27}$ from the underlying QCD theory,
and therefore to gain a dynamical understanding of the so-called
$|\Delta I|=1/2$ rule.
Although this is a very difficult task, considerable progress
has been achieved recently.
Applying the QCD-inspired model of Eq. (\ref{eq:ERTmodel})
to the weak sector,  a quite successful estimate of these
two couplings has been obtained \cite{ref:PR91}.
A very detailed description of
this calculation, and a comparison with other
approaches, can be found in ref. \cite{ref:PR91}.

Once the couplings $g_8$ and $g_{27}$ have been phenomenologically
fixed to the values in Eq. (\ref{eq:g8g27}),
other decays like $K\to 3\pi$ or $K\to 2\pi\gamma$ can be easily
predicted at $O(p^2)$.
As in the strong sector, one reproduces in this way the successful soft-pion
relations of Current Algebra.
However, the data are already accurate enough
for the next-order corrections to be sizeable.
Moreover, many transitions do not occur at $O(p^2)$.
For instance, due to a mismatch between the
minimum number of powers of momenta
required by gauge invariance and the powers of momenta that
the lowest-order
effective Lagrangian can provide,
the amplitude for any non-leptonic radiative
$K$-decay with at most one pion in the final state
($K \rightarrow \gamma
\gamma  , K \rightarrow \gamma l^+ l^- ,
K \rightarrow \pi \gamma \gamma ,
K \rightarrow \pi l^+ l^-$, ...)
vanishes to lowest order in ChPT
\cite{ref:EPR87a,ref:EPR87b,ref:EPR88}.
These decays are then sensitive to the non-trivial quantum field theory
aspects of ChPT.

Unfortunately, at $O(p^4)$ there is a very large number of possible terms,
satisfying the appropriate $(8_L,1_R)$ and $(27_L,1_R)$
transformation properties \cite{ref:KMW90}.
Using the $O(p^2)$ equations of motion obeyed by $U$ to reduce the
number of terms, 35 independent structures
(plus 2 contact terms involving external fields only)
remain in the octet
sector alone \cite{ref:KMW90,ref:EC90}.
Restricting the attention to those terms that contribute to
non-leptonic amplitudes where the only external gauge fields
are photons, still leaves 22 relevant octet terms \cite{ref:EKW93}.
Clearly, the predictive power of a completely general chiral
analysis, using  only symmetry constraints, is rather limited.
Nevertheless, as we are going to see in the next sections,
it is still possible to make predictions.

Due to the complicated interplay of electroweak and strong
interactions, the low-energy constants of the weak non-leptonic
chiral Lagrangian encode a much richer information than
in the pure strong sector.
These chiral couplings contain both long- and short-distance
contributions, and some of them (like $g_8$) have in addition
a CP-violating imaginary part.
Genuine short-distance physics, such as the electroweak penguin
operators, have their corresponding effective realization
in the chiral Lagrangian.
Moreover, there are four $O(p^4)$ terms containing an
$\varepsilon_{\mu\nu\alpha\beta}$ tensor,
which get a direct (probably dominant) contribution from
the chiral anomaly \cite{ref:ENP93,ref:BEP92}.

In recent years, there have been several attempts to estimate
these low-energy
couplings using different approximations, such as
factorization \cite{ref:PR91,ref:CH90},
weak-deformation model \cite{ref:EPR90},
effective-action approach \cite{ref:PR91,ref:BP93},
or resonance exchange \cite{ref:EKW93,ref:IP92}.
Although more work in this direction is certainly needed,
a qualitative picture of the size of the different couplings
is already emerging.

\section{$K\to 2\pi, 3\pi$ decays}
\label{sec:kpp}

Imposing isospin and Bose symmetries, and keeping terms up to
$O(p^4)$, a general parametrization of the $K\to 3\pi$
amplitudes involves ten measurable parameters \cite{ref:DD79},
$\alpha_i$, $\beta_i$, $\zeta_i$, $\xi_i$, $\gamma_3$ and
$\xi'_3$, where $i=1,3$ refers to the $\Delta I =1/2, 3/2$
pieces.
At $O(p^2)$, the quadratic slope parameters
$\zeta_i$, $\xi_i$ and $\xi'_3$ vanish; therefore the
lowest-order Lagrangian (\ref{eq:L2w}) predicts five
$K\to 3\pi$ parameters in terms of the two couplings
$g_8$ and $g_{27}$, extracted from $K\to 2 \pi$.
These predictions give the right qualitative pattern,
but there are sizeable differences with the
measured parameters.
Moreover, non-zero values for some of the slope parameters
have been clearly established experimentally.

The agreement is substantially improved at $O(p^4)$
\cite{ref:KMW91}.
In spite of the large number of unknown couplings in the general
effective $\Delta S=1$ Lagrangian,
only 7 combinations of these weak chiral constants
are relevant for describing the $K\to 2\pi$ and $K\to 3 \pi$
amplitudes \cite{ref:KDHMW92}.
Therefore, one has 7 parameters for 12 observables, which
results in 5 relations.
The extent to which these relations are satisfied provides
a non-trivial test of chiral symmetry at the four-derivative level.
The results of such a test \cite{ref:KDHMW92} are shown in
Table~\ref{tab:KDHMW92}, where the 5 conditions have been
formulated as predictions for the 5 slope parameters.
The comparison is very successful for the two
$\Delta I = 1/2$ parameters.
The data are not good enough to say anything conclusive about
the other three $\Delta I = 3/2$ predictions;
moreover, the possible discrepancy
in the value of $\xi_3$ is not very significative, because
this parameter is expected to be rather sensitive to
electromagnetic effects, which have been omitted in the analysis.

\begin{table}[hbt]
\caption{Predicted and measured values of the quadratic
slope parameters in the $K\to 3\pi$ amplitudes
\protect\cite{ref:KDHMW92}.
All values are given in units of $10^{-8}$.}
\vspace{0.2cm}
\label{tab:KDHMW92}
\begin{center}
\begin{tabular}{|c|c|c|}
\hline
Parameter & Experimental value & Prediction
\\ \hline
$\zeta_1$ & $-0.47\pm0.15$ & $-0.47\pm0.18$ \\
$\xi_1$ & $-1.51\pm0.30$ & $-1.58\pm0.19$ \\
$\zeta_3$ & $-0.21\pm0.08$ & $-0.011\pm0.006$ \\
$\xi_3$ & $-0.12\pm0.17$ & $\hphantom{-}0.092\pm0.030$ \\
$\xi'_3$ & $-0.21\pm0.51$ & $-0.033\pm0.077$ \\ \hline
\end{tabular}
\end{center}
\end{table}

The $O(p^4)$ analysis of these decays has also clarified the
role of long-distance effects ($\pi\pi$ rescattering)
in the dynamical enhancement of $\Delta I = 1/2$ amplitudes.
The $O(p^4)$ corrections give indeed a sizeable
constructive contribution, which results \cite{ref:KMW91}
in a fitted value for $|g_8|$ that is about $30\%$ smaller
than the lowest-order determination (\ref{eq:g8g27}).
While this certainly goes in the right direction,
it also shows that the bulk of the enhancement mechanism
comes from a different source.

 \goodbreak

\section{Radiative $K$ Decays}
\label{sec:radiative}

Owing to the constraints of electromagnetic gauge invariance, radiative
$K$ decays with at most one pion in the final state do not occur
at $O(p^2)$ \cite{ref:EPR87a,ref:EPR87b,ref:EPR88}.
Moreover, only a few terms of the $O(p^4)$ Lagrangian
are relevant for these kinds of processess
\cite{ref:EPR87a,ref:EPR87b,ref:EPR88}:
\beqn
\cL_4^{\Delta S=1,\mbox{\rms em}} & \doteq &
- {G_F\over\sqrt{2}} V_{ud}^{\hphantom{*}}
V_{us}^* \, g_8 \, \biggl\{
-{i e  \over  f^2} F^{\mu \nu}\,
   \left\{ w_1 \,\langle Q \lambda L_\mu L_\nu\rangle
 + w_2 \,\langle Q L_\mu \lambda L_\nu\rangle\right\} \qquad
\biggr.\no\\ & & \biggl. \qquad\qquad\qquad\,\,\,
+  e^2 f^2  w_4 F^{\mu\nu} F_{\mu\nu}\,
 \langle\lambda Q U^\dagger Q U\rangle + \mbox{\rm h.c.} \biggr\}.
\eeqn
The small number of unknown chiral couplings allows us to
derive useful relations among different processes and
to obtain definite predictions. Moreover, the absence of a tree-level
$O(p^2)$ contribution makes the final results very sensitive to the
loop structure of the amplitudes.

\subsection{$K_S\to\gamma\gamma$}

\begin{figure}[h]
\begin{center}
\mbox{\epsfig{file=ksgg.eps,height=7.0cm}}
\end{center}
\caption{Feynman diagrams for $K_1^0\to\gamma^*\gamma^*$.}
\label{fig:ks_gg}
\end{figure}

The symmetry constraints do not allow any direct tree-level
$K_1^0\gamma\gamma$ coupling at $O(p^4)$
($K^0_{1,2}$ refer to the CP-even and CP-odd eigenstates, respectively).
This decay proceeds then
through one loop of charged pions as shown in
Fig.~\ref{fig:ks_gg} (there are
similar diagrams with charged kaons in the loop, but
their sum gives a zero contribution
to the decay amplitude).
Moreover, since there are no possible counter-terms to renormalize
divergences, the one-loop amplitude is necessarily finite.
Although each of the four diagrams in Fig.~\ref{fig:ks_gg}
is quadratically divergent, these divergences cancel  in the
sum.
The resulting prediction \cite{ref:AEG86} is in very good agreement
with the experimental measurement \cite{ref:NA31_87}
\be
Br(K_S\to\gamma\gamma) \, = \,\left\{
\begin{array}{cc}
2.0 \times 10^{-6}   &  \mbox{(theory)}
\\
(2.4 \pm 1.2) \times 10^{-6} & \mbox{(experiment)}
\end{array} \right. .
\ee

\subsection{$K_{L,S}\to\mu^+\mu^-$}

There are well-known short-distance contributions
(electroweak penguins and box diagrams)
to the decay $K_L\to\mu^+\mu^-$.
However, this transition is dominated by long-distance
physics. The main contribution proceeds through a two-photon
intermediate state: $K_2^0\to\gamma^*\gamma^*\to\mu^+\mu^-$.
Contrary to $K_1^0\to\gamma\gamma$,
the prediction for the $K_2^0\to\gamma\gamma$ decay is
very uncertain, because the first non-zero contribution
occurs\footnote{
At $O(p^4)$, this decay proceeds through
a tree-level $K_2^0\to\pi^0,\eta$ transition, followed by
$\pi^0,\eta\to\gamma\gamma$ vertices.
Because of the Gell-Mann--Okubo relation,
the sum of the $\pi^0$ and $\eta$ contributions
cancels exactly to lowest order.
The decay amplitude is then very sensitive to $SU(3)$ breaking.}
at $O(p^6)$.
That makes very difficult any attempt to
predict the $K_{L}\to\mu^+\mu^-$ amplitude.

\begin{figure}[htb]
\begin{center}
\mbox{\epsfig{file=ksmm.eps,height=4.0cm,width=9cm}}
\end{center}
\caption{Feynman diagram for the $K_1^0\to \mu^+ \mu^-$ decay.
The $K_1^0 \gamma^* \gamma^*$ vertex is generated through the one-loop
diagrams shown in Fig.~\protect\ref{fig:ks_gg}}
\end{figure}

The situation is completely different for the $K_S$ decay.
A straightforward chiral analysis \cite{ref:EP91}
shows that, at lowest order in momenta, the only allowed
tree-level $K^0\mu^+\mu^-$ coupling corresponds to the
CP-odd state $K_2^0$.
Therefore, the $K_1^0\to\mu^+\mu^-$ transition can only be
generated by a finite non-local loop contribution.
The two-loop calculation has been performed recently
\cite{ref:EP91}, with the result:
\be
{\Gamma(K_S\to\mu^+\mu^-)\over\Gamma(K_S\to\gamma\gamma)}
= 1.9\times 10^{-6}, \qquad \qquad
{\Gamma(K_S\to e^+ e^-)\over\Gamma(K_S\to\gamma\gamma)}
= 7.9\times 10^{-9},
\ee
well below the present experimental upper limits
\cite{ref:PDG92}.
Although, in view of
the smallness of the predicted ratios,
this calculation seems quite academic, it has important
implications for CP-violation studies.

The longitudinal muon polarization $\cP_L$
in the decay $K_L\to\mu^+\mu^-$ is an interesting measure of CP violation.
As for every CP-violating observable in the neutral kaon system,
there are in general two different kinds of contributions to $\cP_L$:
indirect CP violation through the small $K_1^0$ admixture of the $K_L$
($\varepsilon$ effect), and direct CP violation in the $K_2^0\to\mu^+\mu^-$
decay amplitude.

   In the Standard Model, the direct-CP-violating amplitude is
induced by Higgs exchange with an effective one-loop flavour-changing
$\bar s d H$ coupling \cite{ref:BL86}.
The present lower bound \cite{ref:HIGGS}
on the Higgs mass $m_H>60$ GeV ($95 \%$ C.L.), implies
\cite{ref:BL86,ref:GN89} a
conservative upper limit
$|\cP_{L,\mbox{\rms Direct}}| < 10^{-4}$.
A much larger value $\cP_L \sim O(10^{-2})$ appears quite naturally
in various extensions of the Standard Model  \cite{ref:MO93}.
It is worth emphasizing that $\cP_L$ is especially
sensitive to the presence of light scalars with CP-violating
Yukawa couplings. Thus, $\cP_L$ seems to be a good signature to look
for new physics beyond the Standard Model; for this to be the case,
however, it is very important to have a good quantitative
understanding of the Standard Model prediction to allow us to infer,
from a measurement of $\cP_L$, the existence of a new CP-violation
mechanism.

  The chiral calculation of the $K_1^0\to\mu^+\mu^-$ amplitude
allows us to make a reliable estimate\footnote{
Taking only the absorptive parts of the $K_{1,2}\to\mu^+\mu^-$
amplitudes into account,
a value $|\cP_{L,\varepsilon}| \approx 7\times 10^{-4}$ was
estimated previously \protect\cite{ref:HE83}.
However, this is only one out of four contributions
to $\cP_L$ \protect\cite{ref:EP91},
which
could all interfere constructively with unknown magnitudes.
}
of the contribution to $\cP_L$ due to $K^0$-$\bar K^0$ mixing
\cite{ref:EP91}:
\be\label{eq:p_l}
1.9 < |\cP_{L,\varepsilon}| \times 10^3 \Biggl( {2 \times 10^{-6} \over
Br(K_S\to\gamma\gamma)} \Biggr)^{1/2} < 2.5  .
\ee
Taking into account
the present experimental errors in $Br(K_S\to\gamma\gamma)$  and
the inherent theoretical uncertainties due to uncalculated
higher-order corrections,
one can conclude that experimental indications for
$|\cP_L|>5\times 10^{-3}$ would constitute clear evidence
for additional
mechanisms of CP violation beyond the Standard Model.

\subsection{$K_L\to\pi^0\gamma\gamma$}

Assuming CP conservation, the most general form of the amplitude for
$K_2^0\to\pi^0\gamma\gamma$ depends on two independent invariant
amplitudes $A$ and $B$ \cite{ref:EPR88},
\beqn
\lefteqn{{\cal A}[K_L(p_K)\to \pi^0(p_0)\gamma(q_1)\gamma(q_2)]\, =}
& & \no\\ &&
  \epsilon_\mu(q_1) \,\epsilon_\nu(q_2)
\, \Biggl\{
{A(y,z)\over M^2_K}\,
\Bigl(  q_2^\mu q_1^\nu - q_1\cdot q_2 \, g^{\mu\nu}\Bigr) \,
+ \, {2 B(y,z)\over M^4_K}\,
\Bigl(p_K\cdot q_1 \, q_2^\mu p_K^\nu
\qquad\qquad\Biggr.\Bigr.\no\\\Biggl.\Bigl. &&\qquad\qquad\qquad\quad
+~p_K\cdot q_2\, q_1^\nu p_K^\mu
- p_K^\mu p_K^\nu \, q_1\cdot q_2 -
p_K\cdot q_1\, p_K\cdot q_2 \, g^{\mu\nu}\Bigr)
 \Biggr\} , \qquad
\eeqn
where $y\equiv|p_K\cdot(q_1-q_2)|/M_K^2$ and $z=(q_1+q_2)^2/M_K^2$.

\begin{figure}[thb]
\vfill
\vspace*{-1.0cm}
\begin{minipage}{.47\linewidth}
\begin{minipage}[h]{1.0\textwidth}
\begin{center}
\mbox{\epsfig{file=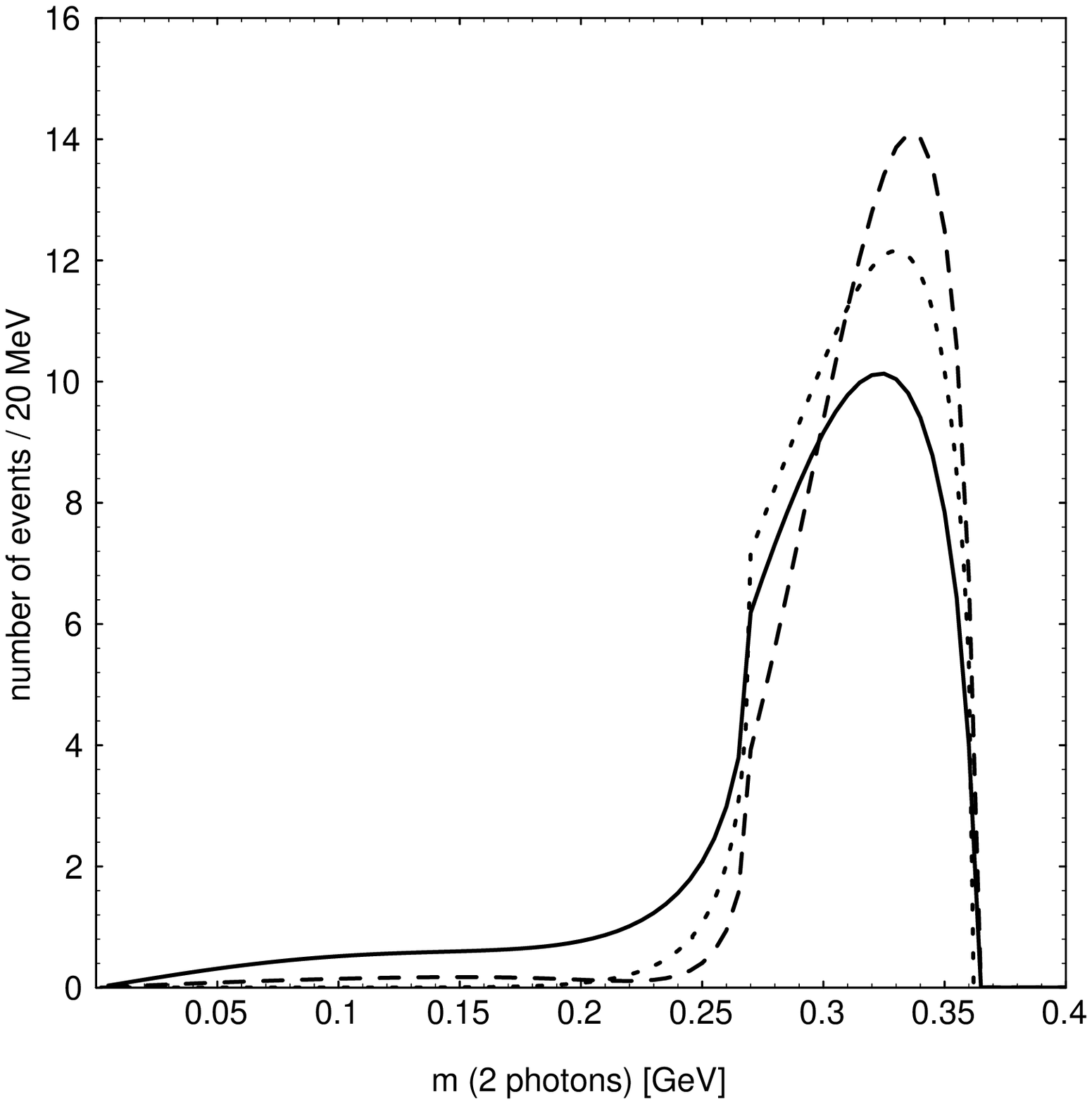,height=10.9cm,width=5.65cm}}
\vspace*{-1.6cm}
\caption{$2\gamma$-invariant-mass distribution for
$K_L\to\pi^0\gamma\gamma$:
$O(p^4)$ (dotted curve),
$O(p^6)$ with $a_V=0$ (dashed curve),
$O(p^6)$ with $a_V=-0.9$ (full curve).
The spectrum is normalized to the 50 unambiguous
events of NA31 (without acceptance corrections).}
\label{fig:spectrum}
\end{center}
\end{minipage}
\end{minipage}
\hspace{0.6cm}
\begin{minipage}{.47\linewidth}
\begin{minipage}[h]{1.0\textwidth}
\begin{center}
\vspace*{2.35 cm}
\mbox{\epsfig{file=NA31.eps,height=6.95cm,width=6.8cm}}
\vspace*{-0.50cm}
\caption{Measured
\protect\cite{ref:NA31_92}
$2\gamma$-invariant-mass distribution for
$K_L\to\pi^0\gamma\gamma$ (solid line).
The dashed line shows the estimated background.
The experimental acceptance is given by the crosses.
The dotted line simulates the $O(p^4)$ ChPT prediction.}
\label{fig:spectrum_NA31}
\end{center}
\end{minipage}
\end{minipage}
\vfill
\end{figure}

Only the amplitude $A(y,z)$ is non-vanishing to lowest non-trivial
order, $O(p^4)$, in ChPT.
Again, the symmetry constraints do not allow any
tree-level contribution from $O(p^4)$ terms in the Lagrangian.
The $A(y,z)$ amplitude is therefore determined by a
finite-loop calculation \cite{ref:EPR87b}.
The relevant Feynman diagrams are analogous to the ones in
Fig.~\ref{fig:ks_gg}, but with an additional $\pi^0$ line
emerging from the weak vertex;
charged kaon loops also give a small contribution in this case.
Due to the large absorptive $\pi^+\pi^-$ contribution,
the spectrum in the invariant mass of the two photons
is predicted \cite{ref:EPR87b,ref:CA88}
to have a very characteristic behaviour
(dotted line in Fig.~\ref{fig:spectrum}),
peaked at high values of $m_{\gamma\gamma}$.
The agreement with the measured two-photon distribution
\cite{ref:NA31_92},
shown in Fig.~\ref{fig:spectrum_NA31},
is remarkably good.
However, the $O(p^4)$ prediction \cite{ref:EPR87b,ref:CA88}
for the rate,
$Br(K_L \rightarrow \pi^0 \gamma \gamma) = 0.67\times 10^{-6}$,
is smaller than the experimental value \cite{ref:NA31_92,ref:E731_91}:
\be
Br(K_L \rightarrow \pi^0 \gamma \gamma )
\, = \,\Biggl\{
\begin{array}{cc}
(1.7 \pm 0.3) \times 10^{-6}  & \mbox{NA31 \cite{ref:NA31_92}},\\
(2.2 \pm 1.0) \times 10^{-6}  & \mbox{E731 \cite{ref:E731_91}}.
\Biggr.
\ea
\ee

Since the effect of the amplitude $B(y,z)$ first appears at
$O(p^6)$, one could worry about the size of the next-order
corrections. In fact, a na\"{\i}ve VMD estimate through the decay
chain
$K_L\to\pi^0,\eta,\eta'\to V \gamma\to\pi^0\gamma\gamma$
\cite{ref:SE90}
results in a sizeable contribution to $B(y,z)$
\cite{ref:EPR90},
\beqn
A(y,z)\big|_{\mbox{\rms VMD}} & = & a_V {G_8 M_K^2 \alpha\over\pi}
\left( 3 - z + {M^2_\pi\over M_K^2}\right) ,
\\
B(y,z)\big|_{\mbox{\rms VMD}} & = & -2 a_V {G_8 M_K^2 \alpha\over\pi} ,
\no
\eeqn
with $a_V \approx 0.32$.
However, this type of calculation predicts a photon
spectrum peaked at low values of $m_{\gamma\gamma}$,
in strong disagreement with
experiment.
As first emphasized in ref.~\cite{ref:EPR90},
there are also so-called direct weak contributions
associated with $V$ exchange, which cannot be written as a strong
VMD amplitude with an external weak transition.
Model-dependent estimates of this direct contribution
\cite{ref:EPR90}
suggest a strong cancellation with the
na\"{\i}ve vector-meson-exchange effect, i.e.
$|a_V| < 0.32$;
but the final result is unfortunately quite uncertain.

A detailed calculation of the most important $O(p^6)$ corrections
has been performed recently \cite{ref:CEP93}.
In addition to the VMD contribution, the unitarity corrections
associated with the two-pion intermediate state
(i.e. $K_L\to\pi^0\pi^+\pi^-\to\pi^0\gamma\gamma$) have been
included\footnote{
The charged-pion loop has also been computed in
ref.~\protect\cite{ref:CAM93}.
}.
Figure~\ref{fig:spectrum} shows the resulting photon spectrum
for $a_V=0$ (dashed curve) and $a_V=-0.9$ (full curve).
The predicted branching ratio is:
\be
BR(K_L\to\pi^0\gamma\gamma) \, = \, \left\{
\begin{array}{cl}
0.67\times 10^{-6} , & O(p^4) , \\
0.83\times 10^{-6} , & O(p^6), \, a_V=0 , \\
1.60\times 10^{-6} , & O(p^6), \, a_V=-0.9 .
\ea
\right.
\ee
The unitarity corrections by themselves raise the rate only
moderately. Moreover, they produce an even more pronounced
peaking of the spectrum at large $m_{\gamma\gamma}$, which
tends to ruin the success of the $O(p^4)$ prediction.
The addition of the $V$ exchange contribution restores again
the agreement.
Both the experimental rate and the spectrum
can be simultaneously reproduced with  $a_V = -0.9$.

\subsection{$K\to\pi l^+ l^-$}

In contrast to the previous processes,
the  $O(p^4)$  calculation of $K^+\to\pi^+ l^+ l^-$
and $K_S\to\pi^0 l^+ l^-$ involves a divergent loop,
which is renormalized by the $O(p^4)$ Lagrangian.
The decay amplitudes
can then be written \cite{ref:EPR87a}
as the sum of a calculable loop
contribution plus an unknown combination of chiral couplings,
\beqn
w_+ & = & -{1\over 3} (4\pi)^2 [w_1^r + 2 w_2^r - 12 L_9^r]
  -{1\over 3} \log{\left(M_K M_\pi/\mu^2\right)} ,
\\
w_S & = & -{1\over 3} (4\pi)^2 [w_1^r - w_2^r]
  -{1\over 3} \log{\left(M_K^2/\mu^2\right)} , \no
\eeqn
where $w_+$, $w_S$
refer to the decay of the $K^+$ and
$K_S$ respectively.
These constants are expected to be of order 1 by
na\"{\i}ve power-counting arguments.
The logarithms have been included
to compensate the
renormalization-scale dependence of the chiral couplings,
so that $w_+$, $w_S$
are observable quantities.
If the final amplitudes are required to transform as
octets, then $w_2 = 4 L_9$, implying
$w_S = w_+ + \log{\left(M_\pi/M_K\right)}/3$
\cite{ref:EPR87a}.
It should be emphasized
that this relation goes beyond the usual requirement
of chiral invariance.

The measured $K^+\to\pi^+ e^+ e^-$ decay rate determines
two possible solutions for $w_+$ \cite{ref:EPR87a}.
The same parameter $w_+$ regulates \cite{ref:EPR87a}
the shape of the invariant-mass distribution of the final lepton
pair.
A fit to the recent BNL E777 data \cite{ref:E777}
gives
\be\label{eq:omega}
w_+ = 0.89^{+0.24}_{-0.14} ,
\ee
solving the previous two-fold ambiguity in favour of the positive
solution, as expected from model-dependent
theoretical estimates \cite{ref:EPR90}.
Once $w_+$ has been fixed, one can make
predictions \cite{ref:EPR87a}
for the rates and Dalitz-plot distributions
of the related modes
$K^+\to\pi^+ \mu^+ \mu^-$,
$K_S\to\pi^0 e^+ e^-$ and $K_S\to\pi^0 \mu^+ \mu^-$.

 The rare decay $K_L \rightarrow \pi^0 e^+ e^-$ is an interesting process
in looking for new CP-violating signatures. If CP were an exact symmetry,
only the CP-even state $K_1^0$ could decay via one-photon emission, while
the decay of the CP-odd state $K_2^0$ would proceed through a two-photon
intermediate state and, therefore,
its decay amplitude would be suppressed
by an additional power of $\alpha$.
When CP-violation is taken into account,
however, an $O(\alpha)$ $K_L \rightarrow \pi^0 e^+ e^-$ decay
amplitude is induced, both through the small
$K_1^0$ component of the $K_L$
($\varepsilon$ effect) and through direct CP-violation in the
$K_2^0 \rightarrow \pi^0 e^+ e^-$ transition.
The electromagnetic suppression of the CP-conserving amplitude then
makes it plausible that this decay is
dominated by the CP-violating contributions.

  The short-distance analysis of the product of weak and electromagnetic
currents allows a reliable estimate of the direct CP-violating
$K_2^0 \rightarrow \pi^0 e^+ e^-$ amplitude.
The corresponding branching ratio induced by
this amplitude has been estimated \cite{ref:ddgb} to be around
\be \label{eq:direct}
Br(K_L \rightarrow \pi^0 e^+ e^-)\Big|_{\mbox{\rms Direct}}
\simeq 5 \times 10^{-12} ,
\ee
the exact number depending on the values of $m_t$ and the quark-mixing
angles.

The indirect CP-violating amplitude induced by the $K_1^0$ component of
the $K_L$ is given by the $K_S \rightarrow \pi^0 e^+ e^-$ amplitude
times the CP-mixing parameter $\varepsilon$.
Using the octet relation between $w_+$ and $w_S$,
the determination of the parameter $\omega_+$ in Eq. (\ref{eq:omega})
implies
\be
Br(K_L \rightarrow \pi^0 e^+ e^-)\Big|_{\mbox{\rms Indirect}} \le
           1.6 \times 10^{-12}.
\ee
Comparing this  value   with the one in Eq. (\ref{eq:direct}),
we see that the interesting direct
CP-violating contribution is expected to be bigger than the
indirect one. This is very different from the situation in
$K \rightarrow \pi \pi$, where the contribution due to mixing
completely dominates.

   The present experimental upper bound \cite{ref:OH90}
(90\% C.L.)
\be
Br(K_L \rightarrow \pi^0 e^+ e^-)\Big|_{\mbox{\rms Exp}}
 < 5.5 \times 10^{-9},
\ee
is still far away from the expected Standard Model signal,
but the prospects
for getting the needed sensitivity of around $10^{-12}$ in
the next few years are rather encouraging.
In order to be able to interpret a future experimental measurement of
this decay as a CP-violating signature, it is first necessary, however,
to pin down the actual
size of the two-photon exchange CP-conserving amplitude.

Using the computed  $K_L\to\pi^0\gamma\gamma$ amplitude,
one can estimate the two-photon exchange contribution
to $K_L\to\pi^0e^+e^-$,
by taking the absorptive part due to the two-photon discontinuity as an
educated guess of the actual size of the complete amplitude.
At $O(p^4)$, the $K_L\to\pi^0e^+e^-$ decay
amplitude is
strongly suppressed (it is proportional to $m_e$), owing to the
helicity structure of the $A(y,z)$ term
\cite{ref:EPR88}:
\be
Br(K_L \rightarrow \pi^0 \gamma ^* \gamma ^* \rightarrow \pi^0
     e^+ e^-)\Big|_{O(p^4)} \,\sim\, 5 \times 10^{-15} .
\ee
This helicity suppression is, however, no longer true at the next order
in the chiral expansion. The $O(p^6)$ estimate of the amplitude
$B(y,z)$ \cite{ref:CEP93} gives rise to
\be
Br(K_L \rightarrow \pi^0 \gamma^* \gamma^* \rightarrow \pi^0 e^+ e^-)
 \Big|_{O(p^6 )} \,\sim\,
\left\{
\begin{array}{cl}
0.3 \times 10^{-12}, & a_V=0 , \\
1.8  \times 10^{-12}, & a_V=-0.9 .
\ea
\right.
\ee
Although the rate increases of course with $|a_V|$,
there is some destructive interference between the unitarity
corrections of $O(p^6)$ and the $V$-exchange contribution
(for $a_V=-0.9$).
In order to get a more accurate estimate,
it would be necessary to make a careful fit to the
$K_L \to\pi^0\gamma\gamma$
data, taking the experimental acceptance into account,
to extract the actual value of $a_V$.

\section{The chiral anomaly in non-leptonic $K$ decays}
\label{sec:anomalous}

The chiral anomaly also  appears in the non-leptonic
weak interactions.
A systematic study of all non-leptonic $K$ decays where
the anomaly contributes at leading order, $O(p^4)$,
has been performed recently
\cite{ref:ENP93}.
Only radiative $K$ decays are sensitive to the
anomaly in the non-leptonic sector.

The manifestations of the anomaly can be grouped in
two different classes of anomalous amplitudes:
reducible and direct contributions.
The reducible amplitudes arise from the contraction of meson
lines between a weak $\Delta S=1$ vertex and the
Wess-Zumino-Witten  functional (\ref{eq:WZW}).
In the octet limit, all reducible anomalous amplitudes of
$O(p^4)$ can be predicted in terms of the coupling $g_8$.
The direct anomalous contributions are generated through
the contraction of the $W$ boson field between
a strong Green function on one side and the
Wess--Zumino--Witten functional
on the other.
Their computation is not straightforward, because of the
presence of strongly interacting fields on both
sides of the $W$.
Nevertheless, due to the non-renormalization theorem
of the chiral anomaly \cite{ref:AB69},
the bosonized form of the direct anomalous amplitudes
can be fully predicted \cite{ref:BEP92}.
In spite of its anomalous origin, this contribution
is chiral-invariant. The anomaly turns out
to contribute to all possible octet terms of
$\cL_4^{\Delta S=1}$ proportional to the
$\varepsilon_{\mu\nu\alpha\beta}$ tensor.
Unfortunately, the coefficients of these terms
get also non-factorizable contributions
of non-anomalous origin, which cannot be computed
in a model-independent way. Therefore, the final
predictions can only be parametrized in terms of four
dimensionless chiral couplings, which are expected
to be positive and of order one.

The most frequent ``anomalous'' decays $K_L\to\pi^+\pi^-\gamma$ and
$K^+\to\pi^+\pi^0\gamma$ share the remarkable feature that the
normally dominant bremsstrahlung amplitude is strongly suppressed,
making the experimental verification of the anomalous amplitude
substantially easier.
This suppression has different origins: $K^+\to\pi^+\pi^0$ proceeds
through the small 27-plet part of the non-leptonic weak interactions,
whereas $K_L\to\pi^+\pi^-$ is CP-violating.
The remaining non-leptonic $K$ decays with direct anomalous contributions
are either suppressed by phase space
[$K^+\to\pi^+\pi^0\pi^0\gamma(\gamma)$,
$K^+\to\pi^+\pi^+\pi^-\gamma(\gamma)$,
$K_L\to\pi^+\pi^-\pi^0\gamma$, $K_S\to\pi^+\pi^-\pi^0\gamma(\gamma)$]
or by the presence
of an
extra photon in the final state
[$K^+\to\pi^+\pi^0\gamma\gamma$, $K_L\to\pi^+\pi^-\gamma\gamma$].
A detailed phenomenological analysis of these decays can be found
in ref.~\cite{ref:ENP93}.

\section{Interactions of a light Higgs}
\label{sec:light_Higgs}

The hadronic couplings of a light Higgs particle are fixed by
low-energy theorems
\cite{ref:GHKD90,ref:CCGBM89,ref:PP90,ref:PPY92},
which relate the $\phi\to\phi' h^0$ transition with
a zero-momentum Higgs to the corresponding $\phi\to\phi'$
coupling.
Although, within the Standard Model, the possibility
of a light Higgs boson is already excluded \cite{ref:HIGGS},
an extended scalar sector with additional degrees of
freedom could easily avoid the present experimental
limits, leaving the question of a light Higgs open to any
speculation.

The quark--Higgs interaction can be written down in the general form
\be
\label{yuk}
\ba
{\cal L}_{h^0 \bar q q} \, = \, - \frac{\dis h^0}{\dis u} \,
\left \{ k_d \, \bar d M_d d \, + \, k_u \, \bar u M_u u \right \},
\ea
\ee
where
$u = (\sqrt{2} G_F)^{-1/2} \approx 246 \,\mbox{\rm GeV}$,
$M_u$ and $M_d$ are the diagonal mass matrices for up-
and down-type quarks respectively, and the couplings $k_u$
and $k_d$ depend on the model considered.
In the Standard Model, $k_u=k_d=1$, while in the usual
two-Higgs-doublet models
(without tree-level flavour-changing neutral currents)
 $k_d = k_u = \cos{\alpha}/\sin{\beta}$ (model I)
or
$k_d = - \sin{\alpha}/\cos{\beta}$,
$k_u = \cos{\alpha}/\sin{\beta}$ (model~II),
where $\alpha$ and $\beta$ are functions of the
parameters of the scalar potential.

The couplings of $h^0$ to the octet of pseudoscalar mesons can be
easily worked out, using ChPT techniques.
The Yukawa interactions of the light-quark flavours can be trivially
incorporated through the external scalar field $s$,
together with the light-quark-mass matrix $\cM$:
\be
s = \cM \left\{ 1 + {h^0\over u} (k_d A + k_u B) \right\},
\ee
where $A\equiv \mbox{\rm diag}(0,1,1)$ and
$B\equiv \mbox{\rm diag}(1,0,0)$.
It remains to compute the contribution from the heavy flavours
$c$, $b$, $t$.
Their Yukawa interactions induce a Higgs--gluon coupling through
heavy-quark loops,
\be
\cL_{h^0GG} = {\alpha_s\over 12\pi} \, (n_d k_d + n_u k_u) \,
{h^0\over u} G^a_{\mu\nu} G_a^{\mu\nu}.
\ee
Here, $n_d=1$ and $n_u=2$ are the number of heavy quarks of
type down and up respectively.
The operator $G^a_{\mu\nu} G_a^{\mu\nu}$ can be related to the
trace of the energy-momentum tensor; in the
three light-flavour theory, one has
\be
\Theta^\mu_\mu = -{b\alpha_s\over 8 \pi} G^a_{\mu\nu} G_a^{\mu\nu}
+ \bar q \cM q ,
\ee
where $b=9$ is the first coefficient of the QCD $\beta$-function.
To obtain the low-energy representation of
$\cL_{h^0GG}$ it therefore suffices to replace
$\Theta^\mu_\mu$
and $\bar q \cM q$ by their corresponding expressions
in the effective chiral Lagrangian theory. One gets
\cite{ref:GHKD90,ref:CCGBM89,ref:PP90},
\be
\cL_{h^0GG}^{\mbox{\rms eff}} = \xi {h^0\over u} {f^2\over 2}\,
\left\{ \langle D_\mu U^\dagger D^\mu U \rangle
+ 3 B_0 \langle U^\dagger \cM + \cM U \rangle \right\} .
\ee
The information on the heavy quarks, which survives in the
low-energy limit, is contained in the coefficient
$\xi \equiv 2 (n_d k_d + n_u k_u)/(3b) = 2 (k_d + 2 k_u)/27$.

Using the chiral formalism, the present experimental
constraints on a very light neutral scalar have
been investigated in refs. \cite{ref:PP90}
and \cite{ref:PPY92}, in the context of two-Higgs-doublet
models.
A Higgs in the mass range $2 m_\mu < m_{h^0} < 2 M_\pi$
can be excluded (within model II), analysing
the decay $\eta\to\pi^0 h^0$ \cite{ref:PP90}.
A more general analysis \cite{ref:PPY92}, using the
light-Higgs production channels
$Z\to Z^*h^0$, $\eta'\to\eta h^0$, $\eta\to\pi^0 h^0$
and $\pi\to e\nu h^0$,
allows us to exclude a large area in the parameter space
($\alpha,\beta,m_{h^0}$) of both models
(I and II) for $m_{h^0} < 2 m_\mu$.

\section{Effective theory at the electroweak scale}
\label{ref:electroweak}

In spite of the spectacular success of the Standard Model, we
still do not really understand the dynamics
underlying the electroweak symmetry breaking
$SU(2)_L\otimes U(1)_Y\to U(1)_{\mbox{\rms em}}$.
The Higgs mechanism provides a renormalizable way
to generate the $W$ and $Z$ masses and, therefore, their
longitudinal degrees of freedom.
However, an experimental verification of this mechanism
is still lacking.

The scalar sector of the Standard Model Lagrangian
can be written in the form
\be
\cL(\Phi) = {1\over 2} \langle D^\mu\Sigma^\dagger D_\mu\Sigma\rangle
- {\lambda\over 16} \left(\langle\Sigma^\dagger\Sigma\rangle
- u^2\right)^2 ,
\ee
where
\be
\Sigma \equiv \left(
\begin{array}{cc}
\Phi^0 & \Phi^+ \\ \Phi^- & \Phi^{0*}
\ea
\right)
\ee
and $D_\mu\Sigma$ is the usual gauge-covariant derivative
\be
D_\mu\Sigma \equiv \partial_\mu\Sigma
+ i g {\vec\tau\over 2} \stackrel{\rightarrow}{W}_\mu \Sigma
- i g' \Sigma {\tau_3\over 2} B_\mu .
\ee
In the limit where the coupling $g'$ is neglected,
$\cL(\Phi)$ is invariant
under global $G\equiv SU(2)_L\otimes SU(2)_C$ transformations,
\be
\Sigma \, \stackrel{G}{\longrightarrow} \,
g_L \,\Sigma\, g_C^\dagger , \qquad\qquad
g_{L,C}  \in SU(2)_{L,C}
\ee
($SU(2)_C$ is the so-called custodial-symmetry group).
The symmetry properties of $\cL(\Phi)$ are very similar to the
ones of the linear-sigma-model Lagrangian (\ref{eq:sigma3}).
Performing an analogous polar decomposition
[see Eqs.~(\ref{eq:polar})],
\beqn
&&\Sigma(x) \, = \, {1\over\sqrt{2}}
\left( u + H(x) \right) \, U(\phi(x)) , \\
&&U(\phi(x)) \, = \, \exp{\left( i \vec{\tau} \vec{\phi}(x) / u \right) } ,
\no
\eeqn
in terms of the Higgs field $H$ and the Goldstones
$\vec{\phi}$,
and taking the limit $\lambda>>1$ (heavy Higgs),
we can rewrite $\cL(\Phi)$ in the standard chiral form:
\be\label{eq:sm_goldstones}
\cL(\Phi) = {u^2\over 4}
\langle D_\mu U^\dagger D^\mu U \rangle  + O\left({H}\right) .
\ee
In the unitary gauge $U=1$, this $O(p^2)$ Lagrangian
reduces to the usual bilinear gauge-mass term.

As we know already, (\ref{eq:sm_goldstones})
is the universal model-independent interaction of the
Goldstone bosons induced by the assumed pattern of SCSB,
$SU(2)_L\otimes SU(2)_C\longrightarrow SU(2)_{L+C}$.
The scattering of electroweak Goldstone bosons
(or equivalently longitudinal gauge bosons)
is then described by the same formulae as
the scattering of pions, changing $f$ by $u$
\cite{ref:CG85}.
To the extent that the present data are still not sensitive to
the virtual Higgs effects, we have only tested up to now
the symmetry properties of the scalar sector encoded in
Eq. (\ref{eq:sm_goldstones}).

In order to really prove the particular scalar dynamics
of the Standard Model, we need to test the
model-dependent part involving the Higgs field $H$.
If the Higgs turns out to be too heavy to be directly
produced (or if it does not exist at all!),
one could still investigate the higher-order effects
\cite{ref:LO81,ref:DEH91,ref:DR90,ref:HT90,ref:GR91,
ref:GE91,ref:FLS91,ref:DV91,ref:RGHM92}
by applying
the standard chiral-expansion techniques in a completely
straightforward way.
The Standard Model gives definite predictions for the
corresponding chiral couplings of the $O(p^4)$ Lagrangian,
which could be tested in future high-precision experiments.
It remains to be seen if the experimental determination
of the higher-order electroweak chiral couplings will confirm
the renormalizable Standard Model Lagrangian,
or will constitute an evidence of new physics

\section{Summary}
\label{sec:summary}

ChPT is a powerful tool to study the low-energy interactions
of the pseudoscalar-meson octet.
This effective Lagrangian framework incorporates all the
constraints implied by the chiral symmetry of the
underlying Lagrangian at the quark--gluon level,
allowing for a clear distinction between genuine aspects of
the Standard Model and additional assumptions of variable
credibility, usually related to the problem of long-distance
dynamics.
The low-energy amplitudes of the Standard Model are calculable in
ChPT, except for some coupling constants which are not restricted by
chiral symmetry. These constants reflect our lack of understanding
of the QCD confinement mechanism and must be determined experimentally
for the time being. Further progress in QCD can only improve
our knowledge of these chiral constants,
but it cannot modify the low-energy structure of the amplitudes.

ChPT provides a convenient language to improve our understanding of
the long-distance dynamics. Once the chiral couplings are
experimentally known, one can test different dynamical models,
by comparing the predictions that they give for those couplings with
their phenomenologically determined values.
The final goal would be, of course, to derive the low-energy
chiral constants
from the Standard Model Lagrangian itself.
Although this is a very difficult problem,
the recent attempts done in this direction look quite promising.

It is important to emphasize that:
\begin{enumerate}
 \item    ChPT is not a model.
The effective Lagrangian generates the more general S-matrix elements
consistent with analyticity, perturbative unitarity and
the assumed symmetries.
Therefore,  ChPT is the effective theory of the Standard Model at
low energies.
\item  The experimental verification of the ChPT predictions
does not provide a test of the detailed dynamics of the
Standard Model; only the implications of the underlying symmetries
are being proved.
Any other model with identical chiral-symmetry properties would give
rise to the same low-energy structure.
\item The dynamical information on the underlying fundamental Lagrangian
is encoded in the chiral couplings.
Different short-distance models with identical symmetry properties
will result in the same effective Lagrangian, but with different
values for the low-energy couplings.
In order to actually test the non-trivial low-energy dynamics
of the Standard Model, one needs first to know the
Standard Model predictions for the chiral couplings.
\end{enumerate}

 In these lectures I have presented the basic formalism of ChPT
and some selected phenomenological applications.
There are many more applications of the chiral framework.
Any system which contains Goldstone bosons can be studied
in a similar way.
A discussion of further topics in ChPT can be found in refs.
\cite{ref:BI93,ref:DO92,ref:EC92,ref:GA90,ref:LE91,ref:ME93,
ref:PI91,ref:RA89,
ref:RINBERG88,ref:DOBOGOKO91,ref:GE84,ref:DGH92}.

\end{document}